\documentclass[10pt,aps,twocolumn,preprintnumbers,prd,noshowpacs,nofootinbib,noshowkeys,floatfix,superscriptaddress]{revtex4}
\usepackage[dvips]{graphics,graphicx}
\usepackage[colorlinks=true,linktocpage=true,linkcolor=blue,citecolor=blue]{hyperref}
\usepackage[usenames,dvipsnames]{color}
\usepackage{amsmath, amssymb,oldgerm}
\usepackage{multirow}
\usepackage{xcolor}
\usepackage[normalem]{ulem}  % \sout{old text} for strikeout
\usepackage{braket}
\usepackage{slashed}
\usepackage[mathscr]{eucal}
\definecolor{goethe-blau}{cmyk}{1.0,0.2,0.0,0.4}
\definecolor{hellgrau}{cmyk}{0.04,0.04,0.05,0.02}
\definecolor{sandgrau}{cmyk}{0.12,0.09,0.13,0.0}
\definecolor{dunkelgrau}{cmyk}{0.25,0.25,0.30,0.75}
\definecolor{emo-rot}{cmyk}{0.04,1.0,0.8,0.07}
\definecolor{purple}{cmyk}{0.08,1.0,0.3,0.36}
\definecolor{senfgelb}{cmyk}{0.01,0.25,1.0,0.05}
\definecolor{gruen}{cmyk}{0.62,0.4,0.87,0.09}
\definecolor{magenta}{cmyk}{0.08,0.86,0.12,0.12}
\definecolor{orange}{cmyk}{0.0,0.7,1.0,0.04}
\definecolor{sonnengelb}{cmyk}{0.0,0.12,0.95,0.0}
\definecolor{helles-gruen}{cmyk}{0.4,0.17,0.81,0.07}
\definecolor{lichtblau}{cmyk}{0.8,0.0,0.06,0.04}

\newcommand{\n}{\nonumber \\ \linebreak}
\newcommand{\cs}{\sigma_{22}}

\renewcommand*{\figurename}{Figure}
\renewcommand*{\tablename}{Table}

\begin{document}

%\preprint{APS/123-QED}

\title{Linear stability analysis of Israel-Stewart theory in the case of a
nonzero background charge}% Force line breaks with \\

\author{Julia Sammet}
\affiliation{Institute for Theoretical Physics, Goethe University,
Max-von-Laue-Str.\ 1, D-60438 Frankfurt am Main, Germany}
\author{Markus Mayer}
\affiliation{Institute for Theoretical Physics, Goethe University,
Max-von-Laue-Str.\ 1, D-60438 Frankfurt am Main, Germany}
\author{Dirk H.\ Rischke}%
\affiliation{Institute for Theoretical Physics, Goethe University,
Max-von-Laue-Str.\ 1, D-60438 Frankfurt am Main, Germany}
\affiliation{Helmholtz Research Academy Hesse for FAIR, Campus Riedberg, 
Max-von-Laue-Str.\ 12, D-60438 Frankfurt am Main, Germany}
%

%\collaboration{CRC-TR 211 Strong-interaction matter under extreme conditions}%\noaffiliation

%\affiliation{
%\affiliation{
%\affiliation{%

%\collaboration{SFB CRC 211 BLABLABLABLA}%\noaffiliation

\date{\today}% It is always \today, today,

\begin{abstract}
Linear stability of Israel-Stewart theory in the presence of net-charge diffusion
was investigated in [Phys.~Rev.~D 102 (2020) 116009] for the case of a massless, classical gas
of noninteracting particles. However, in that work only a vanishing net-charge background was considered. 
In this work, we extend that study to the case of a nonvanishing background charge. 
We find that this effectively results in a change of the numeric value of the charge-diffusion coefficient, in a way
that when the background charge goes to infinity, this coefficient can become at most four times its value at zero 
background charge. We also extend the analysis of [Phys.~Rev.~D 102 (2020) 116009] by performing a systematic parameter study in the plane of charge-diffusion coefficient vs.\ the coupling term between shear-stress and net-charge diffusion. In this plane, we identify regions
where the solutions remain stable and causal
and where they become acausal and/or unstable.
\end{abstract}

\maketitle

\section{Introduction}

Strong-interaction matter in extreme conditions of temperature and density was present in the
early Universe and still exists in the core of neutron stars. In the laboratory, it can be created in
collisions of heavy atomic nuclei at ultrarelativistic energies. The bulk evolution of matter created in
such collisions can be well described by relativistic dissipative fluid dynamics (for a recent review, 
see Ref.\ \cite{Shen:2020mgh}, and refs.\ therein). However, the straightforward relativistic generalization
\cite{Eckart:1940te,1987ifc2}
of nonrelativistic Navier-Stokes theory is acausal and unstable \cite{PhysRevD.31.725}.
The physical reason for this is that in Navier-Stokes theory the dissipative currents, i.e., bulk viscous
pressure, net-charge diffusion current, and shear-stress tensor relax instantaneously to the corresponding dissipative
forces, i.e., expansion scalar, gradients of thermal potential, and fluid shear tensor, respectively.
In order to cure this problem, Israel and Stewart developed a so-called transient, or second-order,
theory of relativistic dissipative fluid dynamics \cite{ISRAEL1976310,stewart1977transient,ISRAEL1979341}, which introduces certain time scales
over which the dissipative currents relax to the values given by the dissipative forces.
Israel-Stewart theory was shown to be causal and stable 
in the linear regime \cite{Hiscock:1985zz,PhysRevD.35.3723,Olson:1990rzl}, provided these so-called relaxation times are sufficiently large, as
demonstrated in Refs.\ \cite{Denicol:2008,Pu:2009fj}. In these works, the causality and stability of Israel-Stewart
theory was analyzed in the case of bulk and shear stress, neglecting the effects of net-charge diffusion. 
Including net-charge diffusion, linear stability and causality of Israel-Stewart theory was addressed in
Ref.\ \cite{Brito:2020nou} for the case of a massless, classical gas
of noninteracting particles, however, only for a vanishing net-charge background.

In this work, we extend the analysis of Ref.\ \cite{Brito:2020nou} in two ways: (i) we
consider the case of nonvanishing net-charge background, and (ii) we perform a more systematic study 
of causality and stability in the plane of net-charge diffusion coefficient vs.\ coupling between shear-stress and
net-charge diffusion.

This work is organized as follows.
In Sec.\ \ref{sec:II} we recall the equations of motion of relativistic second-order dissipative fluid dynamics.
In Sec.\ \ref{sec:III}, these equations are linearized in perturbations on a background which
has constant nonzero energy density, net-charge density, and velocity, respectively. The perturbations of the
fluid-dynamical quantities are tensor-decomposed with respect to the direction of propagation of the perturbation.
We find that a nonzero net-charge background only affects the longitudinal perturbations, while the transverse ones are
the same as for zero net-charge background. 
In Sec.\ \ref{sec:IV} we systematically analyze the dispersion relations of the longitudinal perturbations
in the plane of net-charge diffusion coefficient vs.\ coupling between shear-stress and net-charge diffusion.
We identify various regions in this plane: a region where the system is acausal and thus, in a moving background,
unstable, a region where the system is stable and causal, and a region where the system is unstable but 
remains causal. The latter two regions can be further subdivided according to the specific behavior of the dispersion relations, which
differ qualitatively in these regions. 
We conclude this work with a summary and an outlook in Sec.\ \ref{sec:V}.

We use natural units $\hbar = k_B = c \equiv 1$ and
work in flat Minkowski space with metric tensor $g_{\mu \nu} = \textrm{diag}(+,-,-,-)$.
The spacetime four-vector is denoted as $X^\mu \equiv (t, \mathbf{x})^T$ and
the four-wave number as $K^\mu \equiv (\omega, \mathbf{k})^T$. 

\section{Second-order dissipative fluid dynamics}
\label{sec:II}

The basic equations of fluid dynamics are the conservation laws for charge and 
energy-momentum,
\begin{align}
\partial_\mu N^\mu & = 0\;,\label{netcharge} \\
\partial_{\mu} T^{\mu \nu} &= 0\;.	\label{em}
\end{align}
In the Landau frame, the fluid four-velocity $u^\mu$ follows the flow of energy, such that
the tensor decomposition of the charge current and the energy-momentum tensor
reads
\begin{align}
N^{\mu} &= n u^{\mu} + n^{\mu}\;, \label{nT} \\
T^{\mu \nu} &= \varepsilon u^{\mu} u^{\nu} - (P+\Pi) \Delta^{\mu \nu} + \pi^{\mu \nu}\;.
\label{emT}
\end{align}
Here, $n$ and $\varepsilon$ are the charge density and the energy density in the fluid rest frame, 
$P$ is the pressure, $\Pi$ is the bulk viscous pressure, 
$n^\mu$ is the charge diffusion current and $\pi^{\mu \nu}$
is the shear-stress tensor. Furthermore, $\Delta^{\mu \nu} \equiv g^{\mu \nu}
- u^\mu u^\nu$ is the projector onto the three-space orthogonal to $u^\mu$.
For further use we denote the projection of a four-vector $A^\mu$ onto
this three-space as $A^{\langle \mu \rangle} \equiv \Delta^{\mu}_{\hspace*{0.1cm}\nu} A^\nu$.

Inserting Eqs.\ (\ref{nT}) and (\ref{emT}) into Eqs.\ (\ref{netcharge}) and (\ref{em}), respectively, and
projecting Eq.\ (\ref{em}) onto $u^\mu$ and $\Delta^{\mu \nu}$, respectively, we
arrive at the tensor-projected conservation laws,
\begin{align}
0 & = Dn + n \theta + \nabla_\mu n^\mu - n^\mu Du_\mu\;, \label{ndot} \\
0 & = D\varepsilon + (\varepsilon + P + \Pi) \theta - \pi^{\mu \nu} \sigma_{\mu \nu} \;, 
\label{edot}\\
0 & = (\varepsilon + P + \Pi) Du^\mu - \nabla^\mu (P + \Pi)\n
& \quad + \Delta^\mu_\nu \nabla_\lambda \pi^{\nu \lambda}
- \pi^{\mu \nu} Du_\nu\;. \label{udot}
\end{align}
Here, the comoving derivative of a quantity $A$ is defined as $DA\equiv
u^\mu \partial_\mu A$, while the covariant spatial gradient is denoted as
$\nabla^\mu \equiv \Delta^{\mu \nu} \partial_\nu$. Moreover, $\theta \equiv \partial_\mu
u^\mu$ is the expansion scalar, while $\sigma^{\mu \nu} \equiv
\partial^{\langle \mu} u^{\nu \rangle}$ is
the shear tensor. Here, the tracefree symmetrized projection of a
rank-2 tensor $A^{\mu \nu}$ onto the three-space orthogonal to $u^\mu$ is denoted
as $A^{\langle \mu \nu \rangle} \equiv \Delta^{\mu \nu}_{\alpha \beta} A^{\alpha \beta}$,
where $\Delta^{\mu \nu}_{\alpha \beta} \equiv
\frac{1}{2} (\Delta^\mu_{\hspace*{0.1cm}\alpha} \Delta^\nu_{\hspace*{0.1cm}\beta} + \Delta^\mu_{\hspace*{0.1cm}\beta}\Delta^\nu_{\hspace*{0.1cm}\alpha} - 
\frac{2}{3} \Delta^{\mu \nu} \Delta_{\alpha \beta} )$. 

Equations (\ref{netcharge}) and (\ref{em}) are five equations for fourteen unknowns,
such that we need to provide nine additional equations to close the system of
equations of motion. In second-order dissipative fluid dynamics, these
are relaxation-type equations for the dissipative currents
$\Pi,\, n^\mu$, and $\pi^{\mu \nu}$, which can be derived from an underlying
microscopic theory, e.g., the Boltzmann equation \cite{Denicol:2012cn},
\begin{align}
\tau_\Pi D\Pi + \Pi & = - \zeta \theta - \ell_{\Pi n} \nabla_\mu n^\mu -
\tau_{\Pi n} n^\mu \nabla_\mu P  \n
- &\delta_{\Pi \Pi} \Pi \theta - \lambda_{\Pi n} n^\mu \nabla_\mu \alpha 
+ \lambda_{\Pi \pi} \pi^{\mu \nu}
\sigma_{\mu \nu}\;, \label{tauPi} \\
\tau_n Dn^{\langle \mu  \rangle} + n^{\mu} &= \varkappa \nabla^{\mu} \alpha  
- \tau_n n_{\nu} \omega^{\nu\mu} \n
 -& \delta_{nn} n^{\mu} \theta - \ell_{n\Pi} \nabla^\mu \Pi
+ \ell_{n \pi} \Delta^{\mu\nu} \nabla_{\alpha}\pi^{\alpha}_{\nu} \n  
 + &\tau_{n\Pi} \Pi \nabla^\mu P 
- \tau_{n \pi} \pi^{\mu\nu} \nabla_{\nu}P - \lambda_{nn} n_{\nu} \sigma^{\mu\nu} \n 
+ &\lambda_{n\Pi} \Pi \nabla^\mu \alpha
- \lambda_{n \pi} \pi^{\mu \nu} \nabla_{\nu} \alpha \;, \label{taun}\\
\tau_{\pi}  D\pi^{\langle \mu \nu \rangle} + \pi^{\mu \nu} &= 2 \eta \sigma^{\mu \nu} 
+ 2 \tau_{\pi} \pi_{\lambda}^{\langle \mu} \omega^{\nu \rangle \lambda}  
- \delta_{\pi \pi} \pi^{\mu \nu} \theta \n 
 - &\tau_{\pi \pi} \pi^{\lambda \langle \mu} \sigma^{\nu \rangle}_{\lambda} 
+ \lambda_{\pi \Pi} \Pi \sigma^{\mu \nu}
- \tau_{\pi n} n^{\langle \mu}\nabla^{\nu \rangle} P \n 
+ &\ell_{\pi n} \nabla^{\langle \mu} n^{\nu \rangle} 
+ \lambda_{\pi n} n^{\langle \mu}\nabla^{\nu \rangle} \alpha\;, \label{taupi}
\end{align}
with $\alpha \equiv \beta \mu$, $\beta = 1/T$,
and $\omega_{\mu \nu} \equiv (\nabla_\mu u_\nu - \nabla_\nu u_\mu)/2$ being
the fluid vorticity.

The first terms on the right-hand sides of Eqs.\ (\ref{tauPi}) -- (\ref{taupi}) are
the Navier-Stokes terms, with the bulk-viscosity coefficient $\zeta$, the
charge-diffusion coefficient $\varkappa$, and the shear-viscosity coefficient $\eta$.
These terms are so-called first-order terms, since they are proportional to
gradients of the primary fluid-dynamical variables
$\alpha$, $\beta$ (or $n$, $\varepsilon$), and $u^\mu$.

The coefficients of the first terms on the left-hand sides of Eqs.\ (\ref{tauPi}) -- (\ref{taupi}) 
are the relaxation times $\tau_\Pi, \, \tau_n$, and $\tau_\pi$ 
for the dissipative currents. These terms are of first order in gradients of
dissipative currents. If the dissipative currents are counted as small quantities,
i.e., being of the same order as gradients, these terms are of second order. 
Similar terms also appear on the right-hand sides
of Eqs.\ (\ref{tauPi}) -- (\ref{taupi}), with coefficients $\ell_{\Pi n}$, $\ell_{n \Pi}$,
$\ell_{n \pi}$, and $\ell_{\pi n}$, respectively. All of these terms will play a role
in our linear stability analysis in the next section.

The other terms appearing in Eqs.\ (\ref{tauPi}) -- (\ref{taupi}) are formally also
of second order in small quantities, since they involve the products of
a dissipative current with the gradient of a primary fluid-dynamical quantity.
However, they are nonlinear terms in the sense of a linear stability analysis, since
they involve the product of perturbations, and will thus be neglected in the following.
We remark that second-order dissipative fluid dynamics also features
additional second-order terms. These are either of second order in the
gradients of primary fluid-dynamical quantities or of second order in
dissipative currents \cite{Denicol:2012cn,Molnar:2013lta}. These terms have already been neglected in
Eqs.\ (\ref{tauPi}) -- (\ref{taupi}), since, for the same reasons as the other terms
in these equations, they will not play a role in a linear stability analysis.

\section{Linear stability analysis}
\label{sec:III}

In this section, we linearize the fluid-dynamical equations of motion
(\ref{ndot}) -- (\ref{taupi}), i.e., we consider small perturbations around a 
global-equilibrium state, with $\beta_0 \equiv 1/T_{0}$, 
$\alpha_{0}= \beta_0 \mu_0$, and a fluid four-velocity $u_{0}^{\mu }$, which is
time-like and normalized, $u_{0\mu }u_{0}^{\mu }=1$. We also
introduce the projector $\Delta_0^{\mu\nu} \equiv g^{\mu \nu} - u_0^\mu u_0^\nu$ 
onto the three-space orthogonal to $u_0^\mu$, as well as the
comoving derivative $D_0 A \equiv u_0^\mu \partial_\mu A$ and
the covariant spatial gradient $\nabla_0^\mu \equiv \Delta_{0 \nu}^{\mu} \partial^\nu$
with respect to $u_0^\mu$. Analogously, the tracefree symmetric rank-4 projection
operator orthogonal to $u_0^\mu$ reads
$\Delta^{\mu \nu}_{0\alpha \beta} \equiv
\frac{1}{2} (\Delta^\mu_{0\alpha} \Delta^\nu_{0\beta} + \Delta^\mu_{0\beta} 
\Delta^\nu_{0\alpha} -\frac{2}{3} \Delta_0^{\mu \nu} \Delta_{0\alpha \beta} )$.

The pressure is obtained using an equation of state of the form $P\equiv
P(\alpha,\beta) $. Perturbations $\delta P$ of the pressure are therefore not independent
of the perturbations $\delta \alpha$ and $\delta \beta$. Applying a standard
thermodynamic relation for perturbations around the global-equilibrium background 
leads to
\begin{equation} \label{important_identity}
\beta_0 \delta P = n_0\, \delta \alpha - w_0 \delta \beta\;,
\end{equation}
where $n_{0}\equiv n(\alpha _{0},\beta_0)$, $w_0 \equiv \varepsilon_0 + P_0$, $\varepsilon _{0}\equiv
\varepsilon (\alpha _{0},\beta_0)$, and $P_0 = P(\alpha_0, \beta_0)$ are charge density, 
enthalpy density, energy density, and pressure of the global-equilibrium background.
Choosing $\delta \alpha$ and $\delta \beta$ as the independent perturbations, we
also express the perturbations in energy density and charge density as
\begin{align}
\delta \varepsilon & = \left. \frac{\partial \varepsilon_0}{\partial \alpha_0}
\right|_{\beta_0} \delta \alpha + \left. \frac{\partial \varepsilon_0}{\partial \beta_0}
\right|_{\alpha_0} \delta \beta\;, \\
\delta n& = \left. \frac{\partial n_0}{\partial \alpha_0}
\right|_{\beta_0} \delta \alpha + \left. \frac{\partial n_0}{\partial \beta_0}
\right|_{\alpha_0} \delta \beta\;.
\end{align}

In the global-equilibrium state all dissipative currents vanish,
$\Pi _{0}=\pi _{0}^{\mu\nu }=n_{0}^{\mu }=q_{0}^{\mu }=0$, such that
we obtain for the independent perturbations
\begin{align}
\alpha &=\alpha_{0}+\delta \alpha\; ,  \label{Fluc1} \\
\beta &=\beta_{0}+\delta \beta\; ,  \label{Fluc2} \\
u^{\mu } &=u_{0}^{\mu }+\delta u^{\mu }\; ,  \label{Fluc3} \\
\Pi &=\delta \Pi \;,  \label{Fluc4} \\
n^{\mu } &=\delta n^{\mu }\; ,  \label{Fluc5} \\
\pi ^{\mu \nu } &=\delta \pi ^{\mu \nu }\; ,  \label{Fluc6}
\end{align}
Note that, to first order in perturbations, we also have the relations
$u_0^\mu \delta u_\mu = u_0^\mu \delta n_\mu = u_0^\mu \delta \pi_{\mu \nu} =0$.

Inserting Eqs.\ (\ref{Fluc1}) -- (\ref{Fluc6}) into Eqs.\ (\ref{ndot}) -- (\ref{taupi}),
and neglecting terms of second order in perturbations, we arrive at the
following system of equations of motion,
\begin{align}
0& = \left. \frac{\partial n_0}{\partial \alpha_0}
\right|_{\beta_0} D_0 \delta \alpha + \left. \frac{\partial n_0}{\partial \beta_0}
\right|_{\alpha_0} D_0 \delta \beta \n & \quad + n_0 \nabla_{0\mu} \delta u^\mu 
+ \nabla_{0\mu} \delta n^\mu \;, \label{ndot_2} \\
0 & = \left. \frac{\partial \varepsilon_0}{\partial \alpha_0}
\right|_{\beta_0} D_0 \delta \alpha + \left. \frac{\partial \varepsilon_0}{\partial \beta_0}
\right|_{\alpha_0} D_0 \delta \beta + w_0 
\nabla_{0\mu} \delta u^\mu   \;, 
\label{edot_2} \\
0  & = w_0 D_0 \delta u^\mu - \nabla_0^\mu 
\left(\frac{n_0}{\beta_0} \delta \alpha -
\frac{w_0}{\beta_0} \delta \beta 
 + \delta \Pi\right) \n & \quad + \Delta_{0\nu}^\mu \nabla_{0\lambda} \delta \pi^{\nu \lambda} 
\;, \label{udot_2} 
\end{align}
\begin{align}
0 & = \tau_\Pi D_0 \delta \Pi + \delta \Pi 
+ \zeta \nabla_{0\mu} \delta u^\mu   + \ell_{\Pi n} \nabla_{0\mu} 
\delta n^\mu \;, \label{tauPi_2} \\
0 & = \tau_n \Delta_{0\nu}^\mu D_0 \delta n^{ \nu } + 
\delta n^{\mu} - \varkappa \nabla_0^{\mu} \delta \alpha  \n
& \quad + \ell_{n\Pi} \nabla_0^\mu \delta \Pi
- \ell_{n \pi} \Delta_0^{\mu\nu} \nabla_{0\alpha} \delta \pi^{\alpha}_{\nu}  \;, \label{taun_2}\\
0& = \tau_{\pi}  \Delta^{\mu \nu}_{0 \alpha \beta} D_0
\delta \pi^{\alpha \beta} + \delta  \pi^{\mu \nu} - 2 \eta 
 \Delta^{\mu \nu}_{0 \alpha \beta} \nabla_{0}^\alpha  \delta u^\beta\n
& \quad - \ell_{\pi n} \nabla_0^{\langle \mu} \delta n^{\nu \rangle} \;. \label{taupi_2}
\end{align}

We now solve the system (\ref{ndot_2}) -- (\ref{taupi_2}) of linear partial differential 
equations in Fourier space, i.e., we Fourier-transform the perturbations as
\begin{equation}
\delta A(X) = \int \frac{d^4K}{(2 \pi)^4}\, \delta \tilde{A}(K) \, e^{i K_\mu  X^\mu}\;,
\end{equation}
where $A \in \{\alpha,\beta, u^\mu,\Pi, n^\mu, \pi^{\mu \nu}\}$.
Following Ref.\ \cite{Brito:2020nou} we introduce the quantities
\begin{equation}
\label{eq:27}
\Omega \equiv u_0^\mu K_\mu\;, \quad \kappa^\mu \equiv \Delta_0^{\mu \nu} K_\nu\;.
\end{equation}
Here, $\Omega$ and $\kappa^\mu$ correspond to the frequency and the wave number
of the perturbation in the rest frame of the
background fluid velocity $u_0^\mu$. In Fourier space, 
the system (\ref{ndot_2}) -- (\ref{taupi_2}) 
then reads
\begin{align}
0& =  \Omega \left( \left. \frac{\partial n_0}{\partial \alpha_0}
\right|_{\beta_0}  \delta \tilde{\alpha} + \left. \frac{\partial n_0}{\partial \beta_0}
\right|_{\alpha_0} \delta \tilde{\beta}\right)\n
& \quad +  n_0  \kappa_\mu \,\delta \tilde{u}^\mu 
+  \kappa_\mu \, \delta \tilde{n}^\mu \;, \label{ndot_3} \\
0 & =   \Omega \left(\left. \frac{\partial \varepsilon_0}{\partial \alpha_0}
\right|_{\beta_0}  \delta \tilde{\alpha} + \left. \frac{\partial \varepsilon_0}{\partial \beta_0}
\right|_{\alpha_0}  \delta \tilde{\beta} \right) \n
& \quad +  w_0  \kappa_\mu \,\delta \tilde{u}^\mu   \;, 
\label{edot_3}\\
0  & =  w_0 \Omega \,\delta \tilde{u}^\mu - \kappa^\mu 
\left(\frac{n_0}{\beta_0} \delta \tilde{\alpha} -
\frac{w_0}{\beta_0} \delta \tilde{\beta} 
 + \delta \tilde{\Pi}\right) \n 
& \quad +   \kappa_\nu \, \delta \tilde{\pi}^{\mu \nu} \;, \label{udot_3} \\
0 & =(1+ i \tau_\Pi \Omega )\delta \tilde{\Pi}  
+ i \zeta \kappa_\mu\, \delta \tilde{u}^\mu   + i \ell_{\Pi n} \kappa_\mu \,
\delta \tilde{n}^\mu \;, \label{tauPi_3} \\
0 & = (1+i \tau_n  \Omega )\delta \tilde{n}^{ \mu } 
 - i \varkappa \kappa^{\mu}\, \delta \tilde{\alpha}  \n
& \quad + i \ell_{n\Pi} \kappa^\mu \,\delta \tilde{\Pi}
- i \ell_{n \pi}  \kappa_\nu\, \delta \tilde{\pi}^{\mu \nu}  \;, \label{taun_3}\\
0& = (1+i \tau_{\pi}   \Omega) 
\delta \tilde{\pi}^{\mu\nu}  - 2 i \eta 
 \Delta^{\mu \nu}_{0 \alpha \beta} \kappa^\alpha \, \delta \tilde{u}^\beta\n
& \quad - i \ell_{\pi n} \Delta^{\mu \nu}_{0 \alpha \beta} \kappa_\alpha \,\delta \tilde{n}_\beta\;.
 \label{taupi_3}
\end{align}
Perturbations in the direction of the covariant
wave number $\kappa^\mu$ will decouple from those orthogonal to $\kappa^\mu$.
In order to see this, we now tensor-decompose all quantities with respect to 
$\kappa^\mu$. To this end, we introduce the projection operator
\begin{equation}
\Delta_\kappa^{\mu \nu} \equiv g^{\mu \nu} - \hat{\kappa}^\mu \hat{\kappa}^\nu\;,
\end{equation}
where $\hat{\kappa}^\mu \equiv \kappa^\mu/\kappa$ and $\kappa \equiv 
\sqrt{-\kappa^\mu \kappa_\mu}$ is the modulus of the wave number. 
(Note that Ref.\ \cite{Brito:2020nou} defines $\Delta_\kappa^{\mu \nu}$
with an additional term $-u_0^\mu u_0^\nu$, making it a two-space projector
onto the subspace orthogonal to both $\kappa^\mu$ and $u_0^\mu$.
This is not really necessary, as all vector- and tensor-like perturbations in Eqs.\ (\ref{ndot_3}) -- (\ref{taupi_3})
are already orthogonal to $u_0^\mu$.) Furthermore, the corresponding 
symmetric traceless rank-4 projection operator reads
$\Delta^{\mu \nu}_{\kappa,\alpha \beta} \equiv
\frac{1}{2} (\Delta^\mu_{\kappa, \alpha} \Delta^\nu_{\kappa,\beta} 
+ \Delta^\mu_{\kappa, \beta} 
\Delta^\nu_{\kappa,\alpha} 
-\frac{2}{3} \Delta_\kappa^{\mu \nu} \Delta_{\kappa,\alpha \beta} )$.
A four-vector $A^\mu$ is then decomposed as
\begin{equation}
A^\mu \equiv A_\parallel \hat{\kappa}^\mu + A_\perp^\mu\;, 
\end{equation}
with 
\begin{equation}
A_\parallel \equiv - \hat{\kappa}_\mu A^\mu\;, \quad
A_\perp^\mu \equiv \Delta_\kappa^{\mu \nu} A_\nu\;,
\end{equation}
and the decomposition of a symmetric rank-2 tensor $B^{\mu \nu}$ reads
\begin{align}
B^{\mu \nu} & \equiv B_\parallel \left( \hat{\kappa}^\mu \hat{\kappa}^\nu
+ \frac{1}{3} \Delta_\kappa^{\mu \nu} \right)
+  B_\perp^\mu \hat{\kappa}^\nu
+ B_\perp^\nu \hat{\kappa}^\mu + B_\perp^{\mu \nu}\;, 
\end{align}
with 
\begin{align}
B_\parallel & \equiv B^{\mu \nu} \hat{\kappa}_\mu \hat{\kappa}_\nu\;, \\
B_\perp^\mu & \equiv - \Delta^{\mu \lambda}_\kappa \hat{\kappa}^\nu B_{\lambda \nu}\;, \\
B_\perp^{\mu \nu} & \equiv \Delta_{\kappa,\alpha \beta}^{\mu \nu} B^{\alpha \beta}\;.
\end{align} 
Upon tensor-decomposing Eqs.\ (\ref{ndot_3}) -- (\ref{taupi_3}) with respect
to $\hat{\kappa}^\mu$ we obtain six equations for the perturbations
parallel to $\kappa^\mu$,
\begin{align}
0& =  \Omega \left( \left. \frac{\partial n_0}{\partial \alpha_0}
\right|_{\beta_0}  \delta \tilde{\alpha} + \left. \frac{\partial n_0}{\partial \beta_0}
\right|_{\alpha_0} \delta \tilde{\beta}\right)-  n_0 \kappa  \,\delta \tilde{u}_\parallel 
- \kappa\, \delta \tilde{n}_\parallel \;, \label{ndot_4} \\
0 & =   \Omega \left(\left. \frac{\partial \varepsilon_0}{\partial \alpha_0}
\right|_{\beta_0}  \delta \tilde{\alpha} + \left. \frac{\partial \varepsilon_0}{\partial \beta_0}
\right|_{\alpha_0}  \delta \tilde{\beta} \right) -    
w_0\kappa \,\delta \tilde{u}_\parallel  \;, 
\label{edot_4}\\
0  & =  w_0 \Omega \,\delta \tilde{u}_\parallel - \kappa 
\left(\frac{n_0}{\beta_0} \delta \tilde{\alpha} -
\frac{w_0}{\beta_0} \delta \tilde{\beta} 
 + \delta \tilde{\Pi}\right) \n 
& \quad -  \kappa \, \delta \tilde{\pi}_\parallel\;, \label{udot_4a} \\
0 & = (1+i \tau_\Pi \Omega )\delta \tilde{\Pi}  
- i \zeta \kappa\, \delta \tilde{u}_\parallel   - i \ell_{\Pi n} \kappa \,
\delta \tilde{n}_\parallel \;, \label{tauPi_4}  \\
0 & = (1 + i \tau_n  \Omega )\delta \tilde{n}_\parallel 
 - i \varkappa \kappa\, \delta \tilde{\alpha} 
+ i \ell_{n\Pi} \kappa \,\delta \tilde{\Pi}
+ i \ell_{n \pi}  \kappa\, \delta \tilde{\pi}_\parallel  \;, \label{taun_4a}\\
0& = (1+i \tau_{\pi}   \Omega)\delta \tilde{\pi}_\parallel 
 - \frac{4}{3} i \eta \kappa \, \delta \tilde{u}_\parallel 
  - \frac{2}{3} i \ell_{\pi n} \kappa \,\delta \tilde{n}_\parallel\;,
\label{taupi_4a}
 \end{align}
three equations for the vector-like perturbations orthogonal to $\hat{\kappa}^\mu$,
\begin{align}
0  & =  w_0 \Omega \,\delta \tilde{u}_\perp^\mu -   
\kappa\, \delta \tilde{\pi}^{\mu}_\perp \;, \label{udot_4b} \\
0 & = (1+i \tau_n  \Omega )\delta \tilde{n}^{ \mu }_\perp 
+ i \ell_{n \pi}  \kappa\, \delta \tilde{\pi}^{\mu}_\perp  \;, \label{taun_4b}\\
 0& = (1+i \tau_{\pi}   \Omega) \delta \tilde{\pi}^{\mu}_\perp  -  i \eta 
\kappa \, \delta \tilde{u}^\mu_\perp - i \frac{\ell_{\pi n}}{2}
\kappa \,\delta \tilde{n}^\mu_\perp\;,
\label{taupi_4b}
\end{align}
as well as one equation for the tensor-like perturbation orthogonal
to $\hat{\kappa}^\mu$,
\begin{align}
 0& = (1+i \tau_{\pi}   \Omega) 
\delta \tilde{\pi}_\perp^{\mu\nu}  \n
& \quad -  \frac{i}{9} \kappa \left(2 \eta \,
 \delta \tilde{u}_\parallel +  \ell_{\pi n} \,\delta \tilde{n}_\parallel\right)
 \left( \Delta_\kappa^{\mu \nu} - u_0^\mu u_0^\nu \right)\;.
 \label{taupi_4c}
\end{align}

Again following Ref.\ \cite{Brito:2020nou}, we introduce the time scales
\begin{equation}
\tau_{\eta} \equiv \frac{\eta}{w_0}\;, \quad \tau_\kappa \equiv \frac{\varkappa}{\bar{n}_0}\;,
\quad \tau_\zeta \equiv \frac{\zeta}{w_0}\;,
\end{equation}
where
\begin{equation}
\bar{n}_0 \equiv \frac{\beta_0}{4} w_0\;.
\end{equation}
In the case of a classical ultrarelativistic
gas (i.e., with equation of state $P = \bar{n}T = \varepsilon/3$), $\bar{n}_0$ corresponds to the
total charge density.
We then make all quantities dimensionless, i.e.,
we measure all length and time scales in units of $\tau_\eta$. (Equivalently, we could have also measured them in units of
$\tau_\kappa$ or, in the presence of bulk viscosity, $\tau_\zeta$. However, taking the limit
$\varkappa, \zeta \rightarrow 0$ is then not possible.)
Furthermore, we measure quantities with the dimension of energy density in units
of $w_0$ and quantities with the dimension of density in units of $\bar{n}_0$.
Consequently, the dimensionful
variables are transformed to dimensionless ones as follows,
\begin{align}
\hat{\Omega} & \equiv \tau_{\eta} \Omega \;, \quad \hat{\kappa} \equiv \tau_{\eta} \kappa\;, 
\quad \hat{\tau}_{\Pi,n,\pi, \kappa, \zeta}  \equiv \frac{\tau_{\Pi,n,\pi, \kappa, \zeta}}{\tau_{\eta}}\;, \n
\delta \tilde{X} & \equiv\frac{\delta \tilde{\Pi}}{w_0}\;, \quad \delta \tilde{\xi}^{\mu}  
\equiv \frac{\delta \tilde{n}^{\mu}}{\overline{n}_{0}}\;, 
\quad \delta \tilde{\chi}^{\mu \nu} \equiv \frac{\delta \tilde{\pi}^{\mu \nu}}{w_0}\;, \n
\hat{\mathcal{L}}_{n \pi} & \equiv \frac{4 \ell_{n \pi}}{\beta_0 \tau_{\eta}}\; ,\quad 
\hat{\mathcal{L}}_{\pi n} \equiv \frac{\beta_0 \ell_{\pi n}}{4 \tau_{\eta}}\;, \n
\hat{\mathcal{L}}_{n \Pi} & \equiv \frac{4 \ell_{n \Pi}}{\beta_0 \tau_{\eta}}\; ,\quad 
\hat{\mathcal{L}}_{\Pi n} \equiv \frac{\beta_0 \ell_{\Pi n}}{4 \tau_{\eta}}\;, \label{eq:52}
\end{align}
For dimensionless variables, Eqs.\ (\ref{ndot_4}) -- (\ref{taupi_4a}) can be written in the following matrix form,
\begin{widetext}
\begin{align}
\left ( \begin{array}{cccccc}
\frac{1}{\bar{n}_0} \frac{\partial n_0}{\partial \alpha_0}\hat{\Omega}  &
\frac{\beta_0}{\bar{n}_0} \frac{\partial n_0}{\partial \beta_0} \hat{\Omega} & - \frac{n_{0}}{\bar{n}_{0}} \hat{\kappa} 
& 0 & -\hat{\kappa}  & 0 \\[0.1cm]
\frac{1}{w_0} \frac{\partial \varepsilon_0}{\partial \alpha_0}\hat{\Omega} & 
\frac{\beta_0}{w_0} \frac{\partial \varepsilon_0}{\partial \beta_0} \hat{\Omega}  & -\hat{\kappa}& 0 & 0 & 0\\[0.1cm]
- \frac{n_0}{4\bar{n}_0} \hat{\kappa} &\hat{\kappa}   & \hat{\Omega} & -\hat{\kappa} & 0 & -\hat{\kappa}\\[0.1cm]
0 & 0  & -i \hat{\tau}_\zeta \hat{\kappa}& 1 + i \hat{\tau}_\Pi \hat{\Omega} & - i \hat{\mathcal{L}}_{\Pi n} \hat{\kappa} 
& 0 \\[0.1cm]
- i  \hat{\tau}_{\kappa}\hat{\kappa} &  0 & 0 & i \hat{\mathcal{L}}_{n\Pi} \hat{\kappa} & 1+ i  \hat{\tau}_n  \hat{\Omega}
 &  i\hat{\mathcal{L}}_{n \pi}\hat{\kappa} \\[0.1cm]
  0 & 0 & - \frac{4}{3}i\hat{\kappa} &  0 & - \frac{2}{3}i \hat{\mathcal{L}}_{\pi n}  \hat{\kappa}
  & 1+i   \hat{\tau}_{\pi}  \hat{\Omega}
\end{array}\right)
\left ( \begin{array}{c}
\delta  \tilde{\alpha}\\[0.1cm]
\delta  \tilde{\beta}/\beta_0\\[0.1cm]
\delta   \tilde{u}_\parallel\\[0.1cm]
\delta \tilde{X} \\[0.1cm]
\delta   \tilde{\xi}_\parallel\\[0.1cm]
\delta   \tilde{\chi}_\parallel
\end{array}\right )=\left (\begin{array}{c}
0\\[0.1cm]
0\\[0.1cm]
0\\[0.1cm]
0\\[0.1cm]
0\\[0.1cm]
0
\end{array}\right
) \;.\label{long_modes}
\end{align}
\end{widetext}
After introducing dimensionless variables, Eqs.\ (\ref{udot_4b}) -- (\ref{taupi_4b}) for the transverse fluctuations look 
exactly like Eq.\ (88) in Ref.\ \cite{Brito:2020nou}, i.e., a nonzero background charge does not influence these
modes. We will therefore not consider them further in the following. 
Furthermore, using Eq.\ (\ref{taupi_4a}) one shows that the tensor-like fluctuation, Eq.\ (\ref{taupi_4c}), 
obeys (after introducing dimensionless variables)
\begin{equation}
 \delta \tilde{\chi}_\perp^{\mu \nu} = \frac{1}{6} \delta \tilde{\chi}_\parallel
 \left( \Delta_\kappa^{\mu \nu} - u_0^\mu u_0^\nu \right)\;,
\end{equation}
i.e., it follows the longitudinal fluctuation $\delta \tilde{\chi}_\parallel$. 
Thus, we also do not need to consider this mode any longer. We therefore focus exclusively on the
longitudinal fluctuations in the remainder of this paper.

For the explicit calculation of the longitudinal modes we consider an ideal gas of classical, massless particles, 
i.e., the velocity of sound (squared) is $c_s^2 = 1/3$ and the bulk viscous pressure vanishes. Furthermore,
$w_0 = 4 P_0$, $\bar{n}_0 = P_0 \beta_0$, and from the thermodynamic identity (\ref{important_identity}) and
$n_0 = 2g/(\pi^2 \beta_0^3) \sinh \alpha_0$, where $g$ is the number of internal degrees of freedom, we derive
\begin{align}
\frac{\partial n_0}{\partial \alpha_0} & = \bar{n}_0 \;, \quad \frac{\partial n_0}{\partial \beta_0} = - 3 \frac{n_0}{\beta_0}\;, \\
\frac{\partial \varepsilon_0}{\partial \alpha_0} & = 3 \frac{n_0}{\beta_0} \;, \quad 
\frac{\partial \varepsilon_0}{\partial \beta_0} = - 3 \frac{w_0}{\beta_0}\;.
\end{align}
In this case, Eq.\ (\ref{long_modes}) reduces to
\begin{widetext}
\begin{align}
\left ( \begin{array}{ccccc}
\hat{\Omega}  & - 3\frac{n_0}{\bar{n}_0} \hat{\Omega}  & -\frac{n_{0}}{\bar{n}_{0}}  \hat{\kappa} 
&  -\hat{\kappa}  & 0 \\[0.1cm]
\frac{3 n_0}{4 \bar{n}_0}\hat{\Omega}  & 
-3 \hat{\Omega} & -\hat{\kappa}&  0 & 0\\[0.1cm]
- \frac{n_0}{4\bar{n}_0} \hat{\kappa} &\hat{\kappa}   & \hat{\Omega} &  0 & -\hat{\kappa}\\[0.1cm]
- i  \hat{\tau}_{\kappa}\hat{\kappa} &  0 & 0 &  1+ i\hat{\tau}_n  \hat{\Omega}  
 &  i \hat{\mathcal{L}}_{n \pi}\hat{\kappa}\\[0.1cm]
  0 & 0 & - \frac{4}{3}i\hat{\kappa} &  - \frac{2}{3}i \hat{\mathcal{L}}_{\pi n} \hat{\kappa} 
  & 1+i  \hat{\tau}_{\pi}  \hat{\Omega} 
\end{array}\right)
\left ( \begin{array}{c}
\delta  \tilde{\alpha}\\[0.1cm]
\delta  \tilde{\beta}/\beta_0\\[0.1cm]
\delta   \tilde{u}_\parallel\\[0.1cm]
\delta   \tilde{\xi}_\parallel\\[0.1cm]
\delta   \tilde{\chi}_\parallel
\end{array}\right )=\left (\begin{array}{c}
0\\[0.1cm]
0\\[0.1cm]
0\\[0.1cm]
0\\[0.1cm]
0
\end{array}\right
)\;. \label{long_modes_ur}
\end{align}
\end{widetext}
In order to obtain nontrivial solutions of this linear system of equations one has to require that the determinant of the coefficient matrix vanishes.
This leads to the following condition:
\begin{align}
0 & = \left[ \left( \hat{\Omega}^2 - \frac{\hat{\kappa}^2}{3} \right)\left( 1 + i \hat{\tau}_\pi \hat{\Omega} \right) 
- \frac{4}{3} i \hat{\kappa}^2 \hat{\Omega} \right] \n
& \times \left[ \hat{\Omega} \left( 1 + i \hat{\tau}_n \hat{\Omega} \right) - i \tilde{\tau}_\kappa \hat{\kappa}^2 \right] \n
& - \frac{2}{3} \hat{\mathcal{L}}_{n\pi} \hat{\mathcal{L}}_{\pi n}  \left( \hat{\Omega}^2 - \frac{\hat{\kappa}^2}{3} \right) 
\hat{\kappa}^2 \hat{\Omega}\;, \label{long_modes_2}
\end{align}
where
\begin{equation} \label{tildetaukappa}
\tilde{\tau}_{\kappa} \equiv \hat{\tau}_{\kappa} \left[1- \frac{3}{4} \left(\frac{n_0}{\bar{n}_0}\right)^2\right]^{-1} \;.
\end{equation}
Comparing Eq.\ (\ref{long_modes_2})
with Eq.\ (113) of Ref.~\cite{Brito:2020nou}, we observe that the only effect of a nonvanishing background charge is that the relaxation time 
$\hat{\tau}_\kappa$ is replaced by $\tilde{\tau}_\kappa$.
Varying the background charge $n_0$ 
from 0 to $\pm \bar{n}_0$ (which are the limiting values when $\alpha_0 \rightarrow \pm \infty$), 
the relaxation time $\tilde{\tau}_\kappa$ assumes values from $\hat{\tau}_\kappa$ to $4 \hat{\tau}_\kappa$, i.e., 
it becomes at most four times longer than in the case of zero background charge.

For later purpose, we also write Eq.\ (\ref{long_modes_2}) in the form of Eq.\ (114) of Ref.\ \cite{Brito:2020nou}, 
\begin{align}
& - \mathcal{A} \hat{\Omega}^5 + i \mathcal{B} \hat{\Omega}^4 + (1 + 2 \mathcal{A} \mathcal{S} \hat{\kappa}^2)
\hat{\Omega}^3 \nonumber \\
& - \frac{i}{3} \mathcal{B} \mathcal{D} \hat{\kappa}^2 \hat{\Omega}^2 - \frac{1}{3} ( 1 + \mathcal{E} \hat{\kappa}^2)
\hat{\kappa}^2 \hat{\Omega} + \frac{i}{3} \tilde{\tau}_\kappa \hat{\kappa}^4 = 0\;, \label{eq:60}
\end{align}
where we defined
\begin{align}
\mathcal{A} &\equiv \hat{\tau}_\pi \hat{\tau}_n\;, \\
\mathcal{B} &\equiv \hat{\tau}_\pi + \hat{\tau}_n\;, 
\end{align}
\begin{align}
\mathcal{C} &\equiv \hat{\tau}_n - \frac{1}{2} \hat{\mathcal{L}}_{n\pi } \hat{\mathcal{L}}_{\pi n}\;, \\
\mathcal{D}&\equiv 1 + \frac{3 \tilde{\tau}_\kappa + 4}{\mathcal{B}}\;, \\
\mathcal{E}&\equiv (4 + \hat{\tau}_\pi) \tilde{\tau}_\kappa - \frac{2}{3} \hat{\mathcal{L}}_{n\pi } \hat{\mathcal{L}}_{\pi n}\;, \\
\mathcal{M}&\equiv \frac{\mathcal{E}}{3 \mathcal{A}}\;, \label{def_M} \\
\mathcal{S}&\equiv \frac{\mathcal{A} + 3\hat{\tau}_\pi \tilde{\tau}_\kappa + 4 \mathcal{C}}{6 \mathcal{A}}\;\\
\mathcal{R}&\equiv \sqrt{\mathcal{S}^2 - \mathcal{M}}\;. \label{def_R}
\end{align}
These quantities correspond to those defined in Eqs.\ (90), (91), (115) -- (118) of Ref.\ \cite{Brito:2020nou}, with
the obvious replacement $\hat{\tau}_\kappa \rightarrow \tilde{\tau}_\kappa$.

\section{Results}
\label{sec:IV}

In this section, we discuss the solutions of Eq.\ (\ref{long_modes_2}) or Eq.\ (\ref{eq:60}), respectively. 
These equations contain four different parameters:  $\hat{\tau}_\pi$, $\hat{\tau}_n$, 
$\tilde{\tau}_\kappa$, and $\hat{\mathcal{L}}_{n\pi} \hat{\mathcal{L}}_{\pi n}$. Furthermore, 
one can choose direction and magnitude of the background velocity $u_0^\mu$. 
 
In order to facilitate comparison with the results of Ref.\ \cite{Brito:2020nou}, we fix $\hat{\tau}_\pi = 5$,
$\hat{\tau}_n = 27/4 = 6.75$, and $\hat{\tau}_{\kappa} = 9/16$, corresponding to a constant cross section
in binary scattering of particles. 
Thus, we will study the influence of varying $\tilde{\tau}_\kappa$ (from $\hat{\tau}_{\kappa} = 9/16$
to its maximum value $4\hat{\tau}_{\kappa} = 9/4$) and 
$\hat{\mathcal{L}}_{n\pi} \hat{\mathcal{L}}_{\pi n}$ on the solutions of Eq.\ (\ref{long_modes_2}). 
We note that, for an ultrarelativistic gas of massless particles with constant cross section, 
kinetic theory in the 14-moment approximation predicts $\ell_{n\pi} = \beta_0 \tau_n/20$, $\ell_{\pi n} = 0$
\cite{Denicol:2012cn}, i.e., $\hat{\mathcal{L}}_{n\pi} \hat{\mathcal{L}}_{\pi n}=0$, while a summation of
all moments gives $\ell_{n\pi} \simeq 0.02837\, \beta_0 \tau_n$, $\ell_{\pi n} \simeq -0.56960\, \tau_\pi/\beta_0$
\cite{Wagner:2022ayd}, i.e., $\hat{\mathcal{L}}_{n\pi} \hat{\mathcal{L}}_{\pi n} 
\simeq - 0.01616\, \hat{\tau}_\pi \hat{\tau}_n$. For $\hat{\tau}_\pi = 5$ and $\hat{\tau}_n = 27/4$, we thus have
$\hat{\mathcal{L}}_{n\pi} \hat{\mathcal{L}}_{\pi n}  \simeq - 0.54538$. 

In order to keep the discussion as general
as possible, we will also allow for a nonzero background velocity. Without loss of generality,
we take the three-velocity to point into the $x$-direction, such that $u_0^\mu = \gamma (1, V,0,0)^T$, with
the Lorentz gamma factor $\gamma \equiv (1 - V^2)^{-1/2}$. We only study perturbations travelling in the
same direction, $K^\mu = (\omega, k , 0 , 0)^T$, such that 
\begin{eqnarray}
\hat{\Omega} & = & \gamma ( \hat{\omega} - V \hat{k})\;, \n
- \hat{\kappa}^\mu \hat{\kappa}_\mu \equiv \hat{\kappa}^2 & = & \gamma^2 (\hat{\omega} V - \hat{k})^2\;,
\label{eq:movingomega}
\end{eqnarray}
where we have used Eqs.\ (\ref{eq:27}) and (\ref{eq:52}). 

We first discuss the general structure of the solution to Eq.\ (\ref{eq:60}). Subsequently, we consider 
stability and causality in a background at rest, as well as in a moving background. Deriving analytically
the solution in the limiting cases of zero and infinite wave number allows to deduce 
conditions which delineate the regions
of (in)stability and (a)causality in the $\tilde{\tau}_\kappa - \hat{\mathcal{L}}_{n\pi} \hat{\mathcal{L}}_{\pi n}$ plane.
Finally, we support our findings by showing the numerically computed solutions for selected points in this plane.

\subsection{General structure of solution to Eq.\ (\ref{eq:60})}

After inserting Eq.\ (\ref{eq:movingomega}), Eq.\ (\ref{eq:60}) is a polynomial of order five in $\hat{\omega}$ and
thus has five solutions:  two sound modes, $\hat{\omega}_{s \pm}(\hat{k})$,
one mode associated with charge transport, $\hat{\omega}_{\alpha}(\hat{k})$, 
one mode associated with charge diffusion,
$\hat{\omega}_{n}(\hat{k})$, and one mode associated with shear-stress, $\hat{\omega}_{\pi}(\hat{k})$.
(Note that Ref.\ \cite{Brito:2020nou} chooses a different notation, there the sound modes are
denoted as $\omega_\pm^{\text{sound}}$, while $\hat{\omega}_\alpha \rightarrow \omega^B_{L,-}$,
$\hat{\omega}_n \rightarrow \omega^B_{L,+}$, and $\hat{\omega}_\pi \rightarrow \omega^{\text{shear}}$.)
The first three are hydrodynamic modes, i.e., they vanish in the limit $\hat{k} \rightarrow 0$,
while the latter two are non-hydrodynamic modes, i.e., they assume finite values in the limit $\hat{k} \rightarrow 0$. 

This can be best seen taking the background at rest, $V=0$, for which $\hat{k} \rightarrow \hat{\kappa}$
and $\hat{\omega}_i \rightarrow \hat{\Omega}_i$, cf.\ Eq.\ (\ref{eq:movingomega}), and considering Eq.\ (\ref{eq:60}) for small wave numbers 
$\hat{\kappa} \ll 1$. The five solutions then assume the form
\begin{align}
\hat{\Omega}_{s\pm} (\hat{\kappa}) & = \pm \frac{1}{\sqrt{3}} \hat{\kappa} + \mathcal{O}(\hat{\kappa}^2)\;,  \label{sound}\\
\hat{\Omega}_\alpha (\hat{\kappa})& = i \tilde{\tau}_\kappa \hat{\kappa}^2 + \mathcal{O}(\hat{\kappa}^3)\;, 
\label{charge_transport}\\
\hat{\Omega}_n (\hat{\kappa})& = \frac{i}{\hat{\tau}_n} + \mathcal{O}(\hat{\kappa}^2)\;, \label{charge_diffusion} \\
\hat{\Omega}_\pi (\hat{\kappa})& = \frac{i}{\hat{\tau}_\pi} + \mathcal{O}(\hat{\kappa}^2)\;. \label{shear}
\end{align}
Equations (\ref{sound}) -- (\ref{shear}) will help us to identify the modes when solving Eq.\ (\ref{eq:60})
numerically.

\subsection{Stability and causality in a background at rest}

The stability and causality of the solution of Eq.\ (\ref{eq:60}) in a background at rest was extensively
discussed in Ref.\ \cite{Brito:2020nou}. In essence, our case mirrors their results, except for
the obvious replacement $\hat{\tau}_\kappa \rightarrow \tilde{\tau}_\kappa$.

At $\hat{\kappa} \ll 1$, all modes appear stable, cf.\ Eqs.\ (\ref{sound}) -- (\ref{shear}). However,
Eq.\ (\ref{eq:60}) is of fourth order in the wave number $\hat{\kappa}$,
while in a moving background it is of fifth order in $\hat{k}$ (because for $V>0$ the fifth-order term $\hat{\Omega}^5$ 
now also introduces a fifth-order term $\hat{k}^5$). A moving background
allows to reveal an instability which is hidden for $V=0$.
In order to see this, we will consider the solutions of Eq.\ (\ref{eq:60}) for $\hat{k} =0$ 
in the case of a moving background. This was already done in Ref.\ \cite{Brito:2020nou} 
and we just repeat the discussion for the sake of completeness in Sec.\ \ref{moving}.

Considering the group velocity of the modes in the limit $\hat{\kappa} \rightarrow \infty$ allows to test
causality of the system. Defining
\begin{equation}
\mathcal{T}_\pm \equiv \sqrt{\mathcal{S} \pm \mathcal{R}}
\end{equation}
and inserting the Ansatz $\hat{\Omega} = c \hat{\kappa} + d + \mathcal{O}(\hat{\kappa}^{-1})$, 
one determines the solutions of Eq.\ (\ref{eq:60}) as
\begin{align}
\hat{\Omega}_0(\hat{\kappa})&= i \frac{\tilde{\tau}_\kappa}{\mathcal{E}} + \mathcal{O}(\hat{\kappa}^{-1})\;, 
\label{omega_zero}\\
\hat{\Omega}_{\pm \pm'}(\hat{\kappa})&= \pm' \mathcal{T}_\pm \,\hat{\kappa} + \frac{i}{\mathcal{A}}\,
\frac{3 \mathcal{B} \mathcal{T}_\pm^4 - \mathcal{B} \mathcal{D} \mathcal{T}_\pm^2 + \tilde{\tau}_\kappa}{
15 \mathcal{T}_\pm^4 - 18 \mathcal{S} \mathcal{T}_\pm^2 + 3 \mathcal{M}} \n
& + \mathcal{O}(\hat{\kappa}^{-1})\;,
\label{omegapmpm}
\end{align}
cf.\ Eqs.\ (124) and (125) of Ref.\ \cite{Brito:2020nou}. Here, the index
$\pm \pm'$ stands for the four different combinations $++$, $+-$, $-+$, and $--$, and thus parametrizes four
different modes. Together with the first solution $\hat{\Omega}_0(\hat{\kappa})$ one thus recovers the five independent
solutions of Eq.\ (\ref{eq:60}). It is, however, not immediately obvious how these modes are related to the ones at
$\hat{\kappa} \ll 1$, cf.\ Eqs.\ (\ref{sound}) -- (\ref{shear}). As it will turn out, 
$\hat{\Omega}_{+ \pm} (\hat{\kappa})$ are the two sound modes called $\hat{\Omega}_{s \pm}(\hat{\kappa})$
in Eq.\ (\ref{sound}), while $\hat{\Omega}_{--}(\hat{\kappa})$ is always identical with the charge-diffusion mode
$\hat{\Omega}_n(\hat{\kappa})$ in Eq.\ (\ref{charge_transport}). We will see below that the assignment
of the remaining two modes $\hat{\Omega}_0 (\hat{\kappa})$ and
$\hat{\Omega}_{-+}(\hat{\kappa})$ with the charge-transport mode $\hat{\Omega}_\alpha(\hat{\kappa})$, 
Eq.\ (\ref{charge_diffusion}), and the shear mode $\hat{\Omega}_\pi (\hat{\kappa})$, Eq.\ (\ref{shear}), 
depends on the values of the parameters $\tilde{\tau}_\kappa$
and $\hat{\mathcal{L}}_{n\pi} \hat{\mathcal{L}}_{\pi n}$.

Considering the imaginary parts of the modes in Eqs.\ (\ref{omega_zero}) and (\ref{omegapmpm}), 
the authors of Ref.\ \cite{Brito:2020nou} derived further conditions for the stability of the system. In our case,
the analogous conditions read
\begin{enumerate}
\item[(i)] $\tilde{\tau}_\kappa/\mathcal{E} \geq 0$,
\item[(ii)] $\mathcal{T}_\pm$ are real, and
\item[(iii)] $\pm(3 \mathcal{B} \mathcal{T}_\pm^4 - \mathcal{B} \mathcal{D} \mathcal{T}_\pm^2 + \tilde{\tau}_\kappa)/(
12 \mathcal{R} \mathcal{T}_\pm^2) \geq 0$.
\end{enumerate}
Note that, in order to write condition (iii) in this form, we have further simplified the 
denominator in the second term on the
right-hand side of Eq.\ (\ref{omegapmpm}).
Clearly, since $\mathcal{A} > 0$, $\tilde{\tau}_\kappa >0$, condition (i) is identical to the requirement that
$\mathcal{M} \geq 0$, cf.\ Eq.\ (\ref{def_M}). But then also $\mathcal{R} \leq \mathcal{S}$, cf.\ Eq.\ (\ref{def_R}),
and $\mathcal{T}_-$ is real. We can therefore dispense with condition (i), as it is contained in condition (ii).
The last condition (iii) is equivalent to the requirements
\begin{align}
\mathrm{(iii.1)} & \quad 3 \mathcal{B} \mathcal{T}_+^4 - \mathcal{B} \mathcal{D} \mathcal{T}_+^2 
+ \tilde{\tau}_{\kappa}
 > 0  \;, \label{cond_I} \\
\mathrm{(iii.2)} & \quad 3 \mathcal{B} \mathcal{T}_-^4 - \mathcal{B} \mathcal{D} \mathcal{T}_-^2  
+ \tilde{\tau}_{\kappa}
 < 0  \;.  \label{cond_II}
\end{align}

From Eqs.\ (\ref{omega_zero}), (\ref{omegapmpm}) we observe that, provided $\mathcal{T}_\pm$ is real,
i.e., condition (ii) is fulfilled,
only the modes $\hat{\Omega}_{\pm \pm'}(\hat{\kappa})$ possess 
nonvanishing real parts and thus are propagating modes. 
In this case, the modes $\hat{\Omega}_{+ \pm}(\hat{\kappa})$ are the sound modes 
$\hat{\Omega}_{s \pm}(\hat{\kappa})$, with group velocities 
\begin{equation} 
\lim_{\hat{\kappa} \rightarrow \infty}
 \frac{\partial \mathrm{Re} \hat{\Omega}_{s \pm}(\hat{\kappa})}{\partial \hat{\kappa}} 
= \pm \mathcal{T}_+ \;.\label{eq:agv_sound}
\end{equation}
The modes $\hat{\Omega}_{- \pm}(\hat{\kappa})$ are the charge-transport and charge-diffusion modes 
$\hat{\Omega}_{\alpha, n}(\hat{\kappa})$, 
with group velocities
\begin{equation} \label{eq:agv_shear}
 \lim_{\hat{\kappa} \rightarrow \infty}
 \frac{\partial \mathrm{Re} \hat{\Omega}_{\alpha, n}(\hat{\kappa})}{\partial \hat{\kappa}} 
= \pm \mathcal{T}_-\;.
\end{equation}\\
This leaves $\hat{\Omega}_0(\hat{\kappa}) \equiv \hat{\Omega}_\pi (\hat{\kappa})$ as the nonpropagating mode.
As long as $\mathcal{T}_-$ is real,
$\hat{\Omega}_{-+}(\hat{\kappa})$ corresponds to the charge-transport mode, and not to the shear mode.
This changes for imaginary $\mathcal{T}_-$.
Causality of the system now requires that
\begin{equation}
\mathcal{T}_\pm \leq 1\;,
\end{equation}
cf.\ Eq.\ (126) of Ref.\ \cite{Brito:2020nou}.

\subsection{Causality and stability in a moving background}
\label{moving}

We now generalize the results of the previous subsection to the case $V>0$.
The solutions for $\hat{k} =0$ in a moving background have already been discussed in Ref.\ \cite{Brito:2020nou}.
We just repeat the discussion here for the sake of completeness.
 Considering Eq.\ (\ref{eq:60}) for $\hat{k} = 0$ leads to three solutions 
$\hat{\omega}_{s \pm, \alpha} (\hat{k} = 0) =0$, corresponding to the two sound modes and the charge-transport mode.
This is analogous to the case when the background is at rest, cf.\ Eqs.\ (\ref{sound}) and (\ref{charge_transport}).
The other two solutions correspond to the shear and charge-diffusion modes,
\begin{widetext}
\begin{equation} \label{eq:sol2}
\hat{\omega}_{\pi,n}(\hat{k} = 0) = i \frac{\tilde{\tau}_\kappa V^4 + \mathcal{B} (3- \mathcal{D}V^2)
\pm \sqrt{ \left[ \tilde{\tau}_\kappa V^4 + \mathcal{B} (3- \mathcal{D}V^2)\right]^2 - 12 \mathcal{A} (3 - V^2)
(1 - 2 \mathcal{S}V^2 + \mathcal{M}V^4)}}{6 \gamma \mathcal{A} (1 - 2 \mathcal{S}V^2 + \mathcal{M}V^4)}\;.
\end{equation}
\end{widetext}
Note that there is a factor of 2 missing in the denominator of Eq.\ (132) of Ref.\ \cite{Brito:2020nou}. Other than that,
these solutions are the same, except for the obvious replacement $\hat{\tau}_\kappa \rightarrow \tilde{\tau}_\kappa$.
In the limit $V \rightarrow 0$, Eq.\ (\ref{eq:sol2}) reduces to
\begin{equation} \label{eq:83}
\hat{\omega}_n(\hat{k} = 0) = \frac{i}{\hat{\tau}_n}\;, \quad \hat{\omega}_\pi(\hat{k} = 0)= \frac{i}{\hat{\tau}_\pi}\;,
\end{equation}
as expected, cf.\ Eqs.\ (\ref{charge_diffusion}) and (\ref{shear}). 
For $V >0$, one must have $1- 2 \mathcal{S}V^2 + \mathcal{M}V^4 >0$, otherwise one solution is always unstable.
The equality $1- 2 \mathcal{S}V^2 + \mathcal{M}V^4 =0$ has two roots $V_\pm = 1/\mathcal{T}_\mp$,
and one must ensure that already the smaller one, $V_- = 1/\mathcal{T}_+$, is larger than one, or equivalently
$\mathcal{T}_+ <1$, otherwise there exists
a frame which moves with a velocity $V$ which fulfills $1/\mathcal{T}_+ < V \leq 1$. 
But $\mathcal{T}_+ $ is exactly the modulus of the asymptotic
group velocity of the sound modes, cf.\ Eq.\ (\ref{eq:agv_sound}), 
so this requirement is identical to the requirement that
this asymptotic group velocity remains causal. As already noted in Ref.\ \cite{Brito:2020nou}, the causality
of the solution for $\hat{k} \rightarrow \infty$ implies the stability at $\hat{k} =0$, or vice versa, a violation of causality
would imply an instability.

In addition, one needs to makes sure that $\tilde{\tau}_\kappa V^4 + \mathcal{B} (3- \mathcal{D}V^2) >0$.
The equality $\tilde{\tau}_\kappa V^4 + \mathcal{B} (3- \mathcal{D}V^2) =0$ has two roots,
\begin{equation}
V_\pm^2 = \frac{1}{2 \tilde{\tau}_\kappa} \left( \mathcal{B} \mathcal{D} \pm \sqrt{ \mathcal{B}^2 \mathcal{D}^2 
- 12 \tilde{\tau}_\kappa \mathcal{B}}
\right)\;.
\end{equation} 
Again, one must ensure that the smaller one is outside the physical region, i.e.,
$V^2_- \geq 1$, which leads to the requirement $\hat{\tau}_\pi + \hat{\tau}_n \geq 2 + \tilde{\tau}_\kappa$, cf.\ Eq.\ (139)
of Ref.\ \cite{Brito:2020nou}. For our choice of parameters, this implies
$\tilde{\tau}_\kappa \leq 3 + \frac{27}{4} = 39/4$. Since $\tilde{\tau}_\kappa$ is
restricted to the range $9/16 \leq 
\tilde{\tau}_\kappa \leq 9/4$, this requirement is always fulfilled for the cases studied in this work.

Extending the discussion of Ref.\ \cite{Brito:2020nou} we now study causality in the moving frame.
This requires an analysis of Eq.\ (\ref{eq:60}) in the limit $\hat{k} \rightarrow \infty$. We insert the ansatz
$\hat{\omega} = c \hat{k} + d + \mathcal{O}(1/k)$ into this equation and derive to order $\mathcal{O}(\hat{k}^5)$ the
condition
\begin{align}
 (c-V)  &\left[ \mathcal{A} (c-V)^4 - 2 \mathcal{A} \mathcal{S} (c-V)^2 (1-cV)^2 \right. \n
&  \left. + \frac{\mathcal{E}}{3} (1-cV)^4 \right] =0\;, \label{eq:order5}
\end{align}
while to order $\mathcal{O}(\hat{k}^4)$ we obtain
\begin{widetext}
\begin{align}
& - 5 \mathcal{A} \gamma d (c-V)^4 + i \mathcal{B} (c-V)^4 + 2 \mathcal{A} \mathcal{S} \gamma d (c-V) (1-cV)\left[ 3(c-V) (1-cV)- 2V (c-V)^3 \right] \n
& - \frac{i}{3} \mathcal{B} \mathcal{D} (c-V)^2 (1-cV)^2
- \frac{\mathcal{E}}{3} \gamma d (1-cV)^3 \left[ 1-cV - 4 V (c-V) \right] + \frac{i}{3} \tilde{\tau}_\kappa (1-cV)^4 =0\;.
\label{eq:order4}
\end{align}
\end{widetext}
Using Eq.\ (\ref{def_M}), it is easy to see that Eq.\ (\ref{eq:order5}) has five solutions, 
$c_0 = V$, corresponding to an advected, nonpropagating mode, and
\begin{equation}
\frac{c_{\pm \pm'} -V}{1-c_{\pm \pm'} V} = \pm' \mathcal{T}_\pm\;,
\end{equation}
or
\begin{equation}
c_{\pm \pm'} = \frac{V \pm' \mathcal{T}_\pm}{1 \pm'  V \mathcal{T}_\pm}\;,
\end{equation}
where the notation $\pm'$ again means that the two signs are independent from those of $\pm$.
One readily recognizes the relativistic addition theorem for velocities, which implies that as long as
$1 \geq \mathcal{T}_\pm \in \mathbb{R}$ and $V \leq 1$, one also has $c_{\pm \pm'} \leq 1$. 
This, in turn, has the consequence that the asymptotic group velocity
\begin{equation}
\lim_{\hat{k} \rightarrow \infty} \frac{\partial \mathrm{Re} \,\hat{\omega}_{\pm \pm'}(\hat{k})}{\partial \hat{k}}
= c_{\pm \pm'} \leq 1\;,
\end{equation}
i.e., all modes are causal. But the condition $\mathcal{T}_\pm \leq 1$ is the same causality condition as for
the case $V=0$, cf.\ Eqs.\ (\ref{eq:agv_sound}), (\ref{eq:agv_shear}). We now show that also
the stability conditions (i) -- (iii) are identical to those for $V=0$.

Inserting the solution $c_0$ into Eq.\ (\ref{eq:order4}), we obtain
\begin{equation}
d_0 = i \frac{\tilde{\tau}_\kappa}{\gamma \mathcal{E}}\;,
\end{equation}
i.e., this mode is stable as long as $\tilde{\tau}_\kappa/\mathcal{E} \geq 0$, which is the
same as condition (i) in the case of a background at rest (the additional factor $\gamma \geq 1$ appearing in
the denominator of this equation does not affect this conclusion).
As discussed above, this is equivalent to demanding $\mathcal{T}_- \in \mathbb{R}$,
which is one part of condition (ii), as this implies $\mathcal{S} \geq \mathcal{R}$, i.e., 
according to Eq.\ (\ref{def_R}) $\mathcal{M} \geq 0$, and therefore also $\mathcal{E} \geq 0$,
cf.\ Eq.\ (\ref{def_M}). We will see that, as long as this is fulfilled, the stable mode $\hat{\omega}_0 (\hat{k})$ corresponds
to the nonpropagating shear mode $\hat{\omega}_\pi (\hat{k})$. Once $\mathcal{M} <0$, however, this unstable
mode becomes the (nonpropagating) charge-transport mode $\hat{\omega}_\alpha (\hat{k})$.

\begin{figure*}[!]
	\includegraphics[scale=0.6]{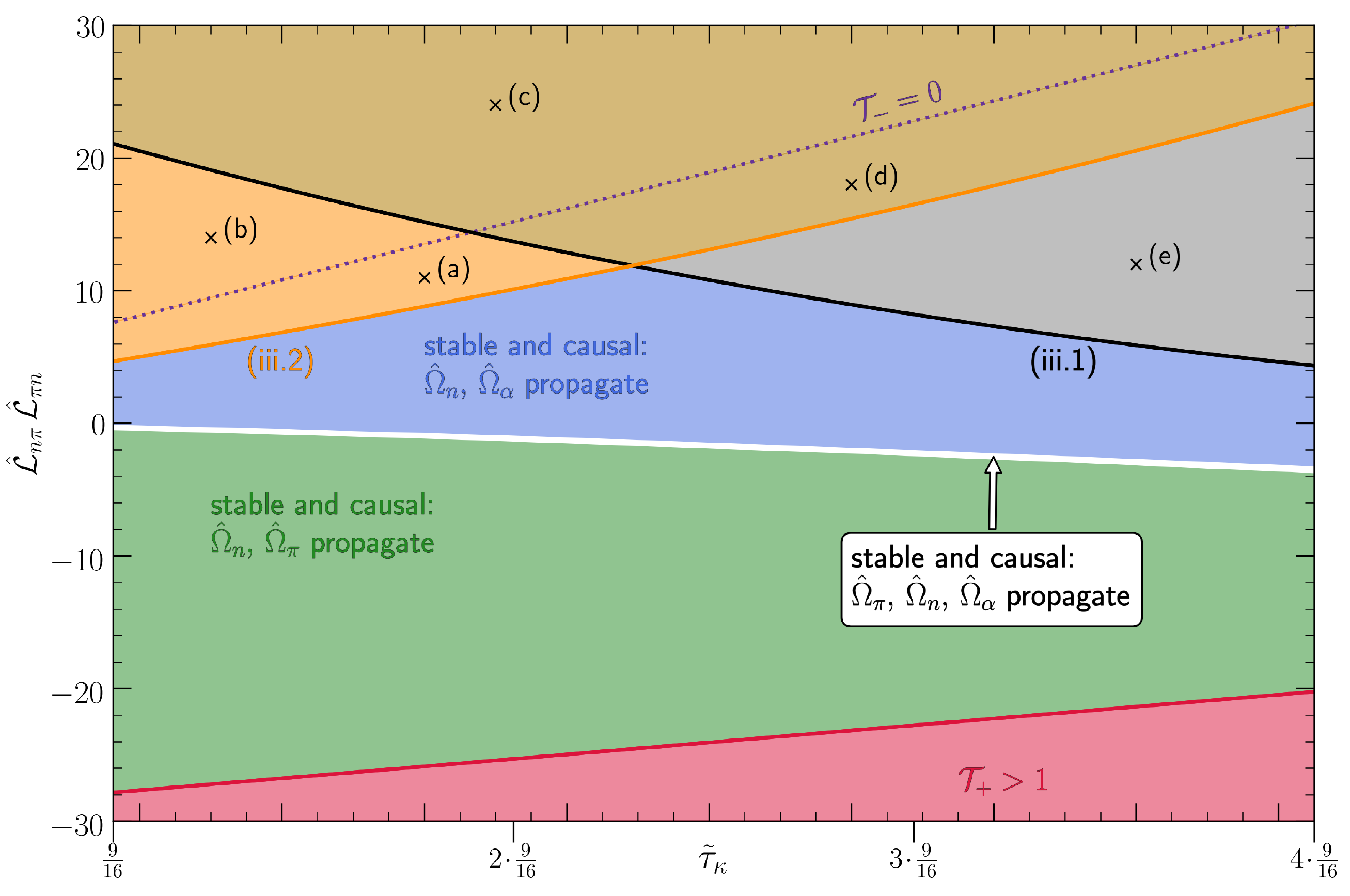}
	 \caption{Different regions of (in-)stability and (a-)causality in the 
	 $\tilde{\tau}_\kappa - \hat{\mathcal{L}}_{n\pi} \hat{\mathcal{L}}_{\pi n}$ plane. For explanations see text.}
	 \label{figure3}
\end{figure*}

Finally, inserting $c_{\pm \pm'}$ into Eq.\ (\ref{eq:order4}) one obtains
\begin{align}
d_{\pm \pm'} & =  \pm \frac{i}{\gamma \mathcal{A}} \, \frac{3 \mathcal{B} \mathcal{T}_\pm^4 - \mathcal{B} \mathcal{D}
\mathcal{T}_\pm^2 + \tilde{\tau}_\kappa}{12 \mathcal{R} \mathcal{T}_\pm^2 \left( 1 \pm' V \mathcal{T}_\pm \right)}\;. \label{eq:91}
\end{align}
The change in the corresponding result for the case $V=0$ (condition (iii) above) is
a factor $\gamma (1 \pm' V \mathcal{T}_\pm)$ in the denominator, which is positive as long as 
$V \leq 1$, $\mathcal{T}_\pm \leq 1$, i.e., as long as
causality of the asymptotic group velocities is maintained. 
For stability, we thus have to demand the same stability conditions (iii.1) and (iii.2) as in the case of a static background,
Eqs.\ (\ref{cond_I}) and (\ref{cond_II}). Vice versa, if one of these conditions are violated, we have explicitly
identified modes with a negative imaginary part, i.e., they become unstable, although their
asymptotic group velocities remain causal. For instance, if condition (iii.1) is violated, we have
$d_{+ \pm} <0$, which means that the two sound modes $\hat{\omega}_{s \pm} (\hat{k})$ become unstable.
On the other hand, violation of condition (iii.2) implies that $d_{-\pm} < 0$. Then, as long as
$\mathcal{M} \geq 0$, the modes 
$\hat{\omega}_{\alpha, n} (\hat{k})$ become unstable. Once $\mathcal{M}<0$, one would
expect that $\hat{\omega}_{\pi, n}(\hat{k})$ become unstable, because
these modes correspond to $\hat{\omega}_{-\pm}(\hat{k})$, since, as mentioned above, the nonpropagating
mode $\hat{\omega}_0 (\hat{k})$ is now the charge-transport mode
$\hat{\omega}_\alpha (\hat{k})$ (which is also unstable). However, the situation is
more subtle, as then $\mathcal{T}_-$ becomes imaginary. We will comment on what happens in this
case in the next paragraph. 

Let us close the discussion of Eq.\ (\ref{eq:91}) by noting that,
once $\mathcal{T}_\pm >1$, there exists a
frame with velocity $V <1$ for which
$V \mathcal{T}_\pm$ can be larger than
1. Even when both conditions (iii.1) and
(iii.2) are fulfilled, this flips the 
sign of the $\pm -$ modes,
i.e., $d_{\pm -}$ can become negative,
signaling that the respective modes become
unstable.

We close this general discussion of stable and unstable modes by noting that $\mathcal{T}_+$ remains 
real for all choices of parameters, no matter whether $\mathcal{M}$ is positive or negative. This
implies that, at least for $\hat{k} \gg 1$, the two sound modes are always propagating modes. However,
when $\mathcal{M} <0$, then, as just mentioned, $\mathcal{T}_-$ becomes imaginary. This means that the
dispersion relation for the corresponding modes, which are the charge-diffusion and the shear modes,
become $\hat{\omega}_{- \pm} (\hat{k})= \pm i |\mathcal{T}_-| \hat{k} + i d_{- \pm} + \mathcal{O}(\hat{k}^{-1})$.
The mode $\hat{\omega}_{-+}(\hat{k})$, which turns out to be the shear mode $\hat{\omega}_{\pi}(\hat{k})$,
is then stable, at least for $|\mathcal{T}_-|\hat{k} + d_{-+} > 0$ (which is always fulfilled for
sufficiently large $\hat{k}$), while
the mode $\hat{\omega}_{--}(\hat{k})$, which is the charge-diffusion mode $\hat{\omega}_{n}(\hat{k})$, is always unstable, 
and even increasingly more so as $\hat{k}$ grows. As mentioned above, the nonpropagating mode is
now $\hat{\omega}_{\alpha}(\hat{k})$, which becomes unstable since $\mathcal{E} < 0$. Thus, ultimately
the charge-transport and charge-diffusion modes are again the unstable modes, just as for $\mathcal{E} \geq 0$.

\begin{figure*}[!]
	 \includegraphics[scale=0.5]{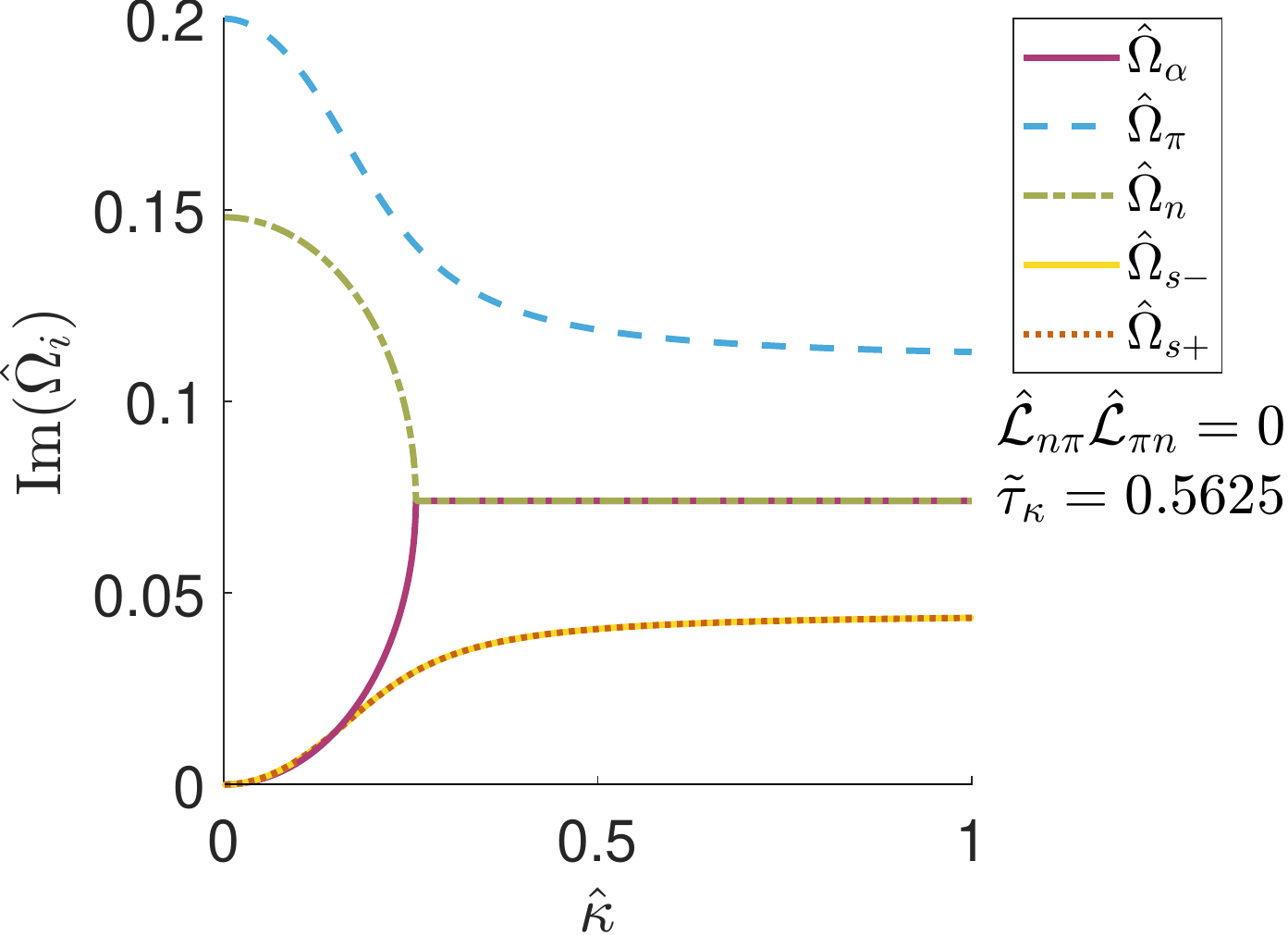} \hspace*{0.5cm}
	 \includegraphics[scale=0.5]{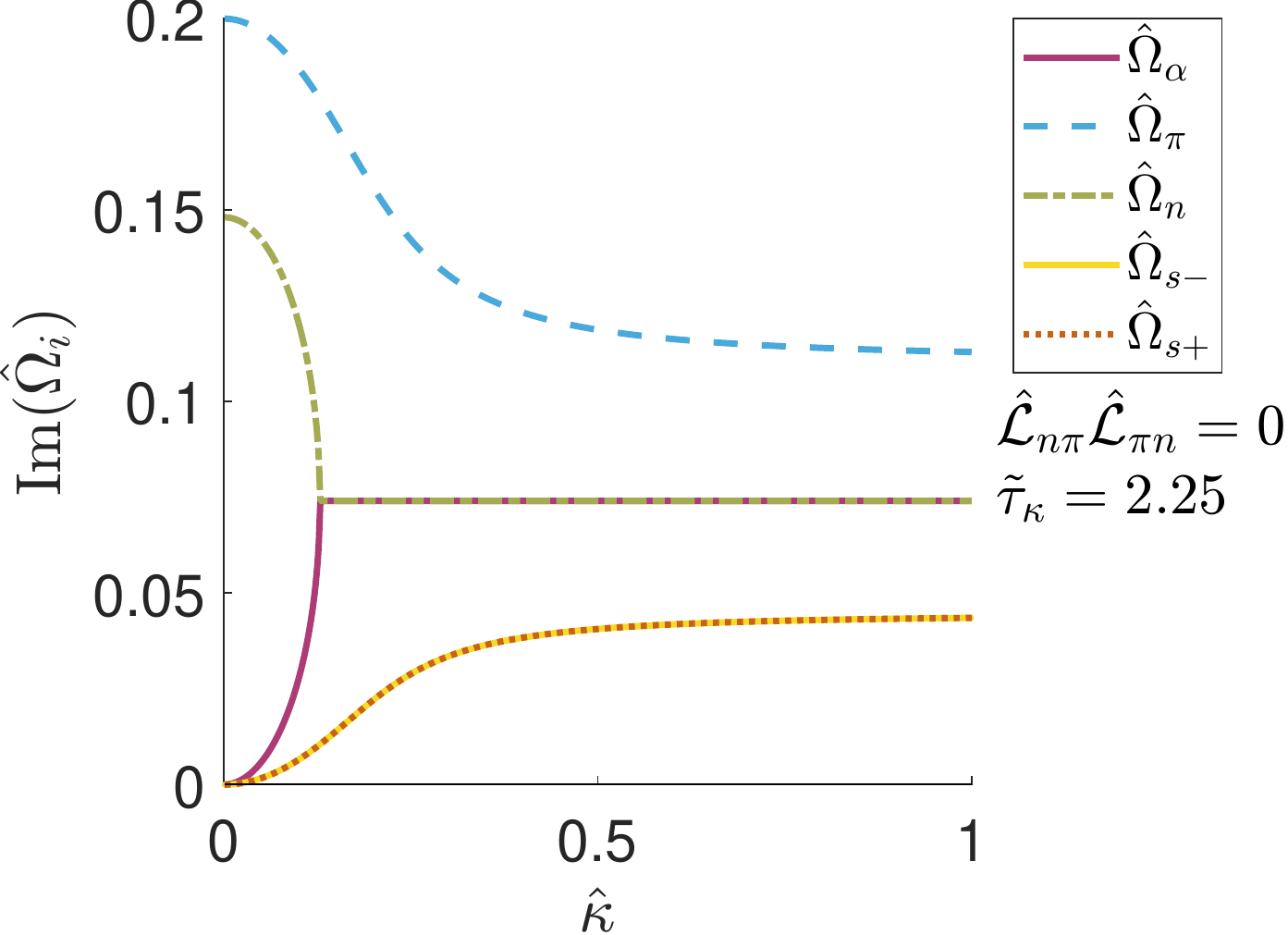}\\
	 \includegraphics[scale=0.5]{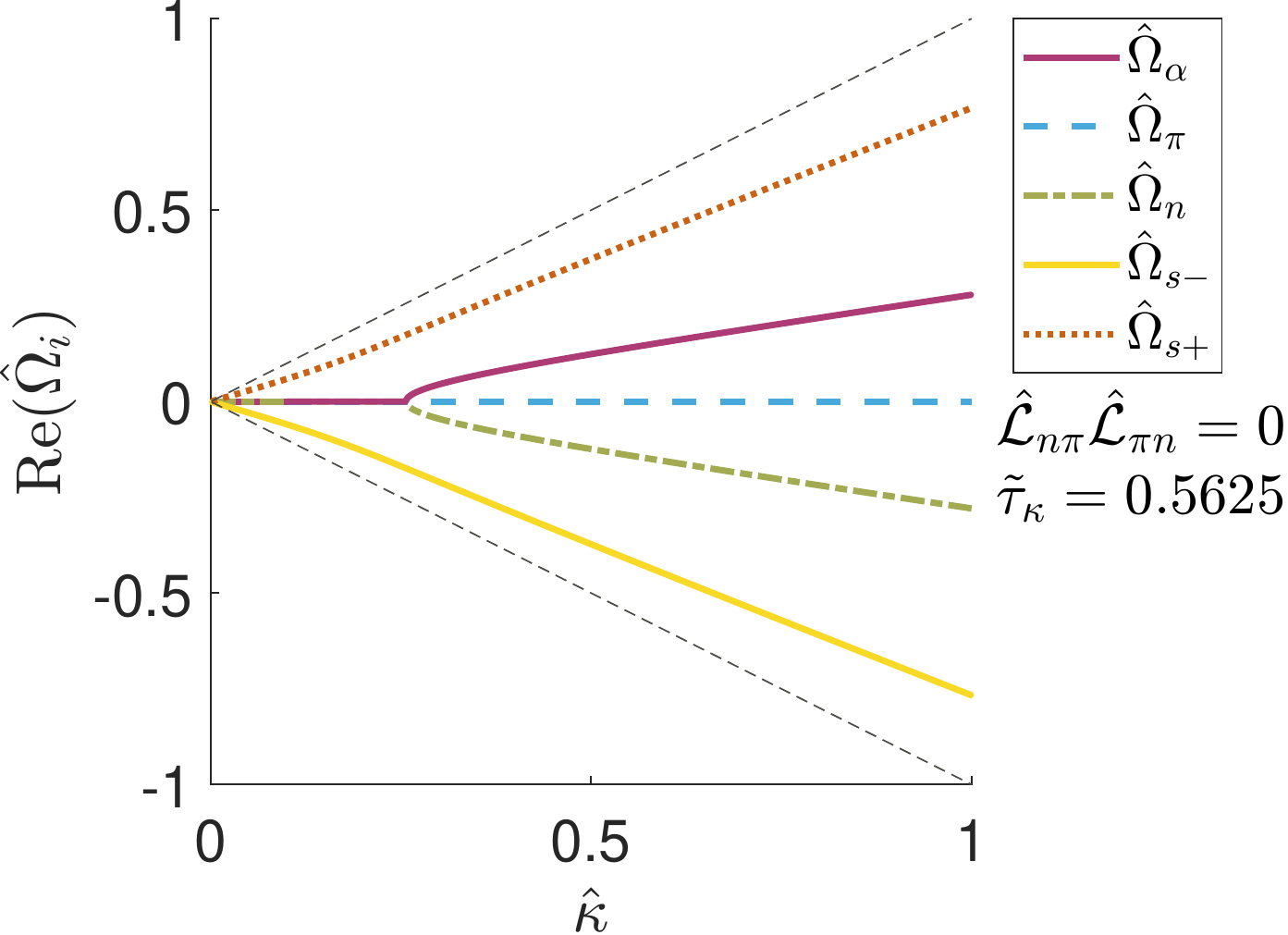} \hspace*{0.5cm}
	 \includegraphics[scale=0.5]{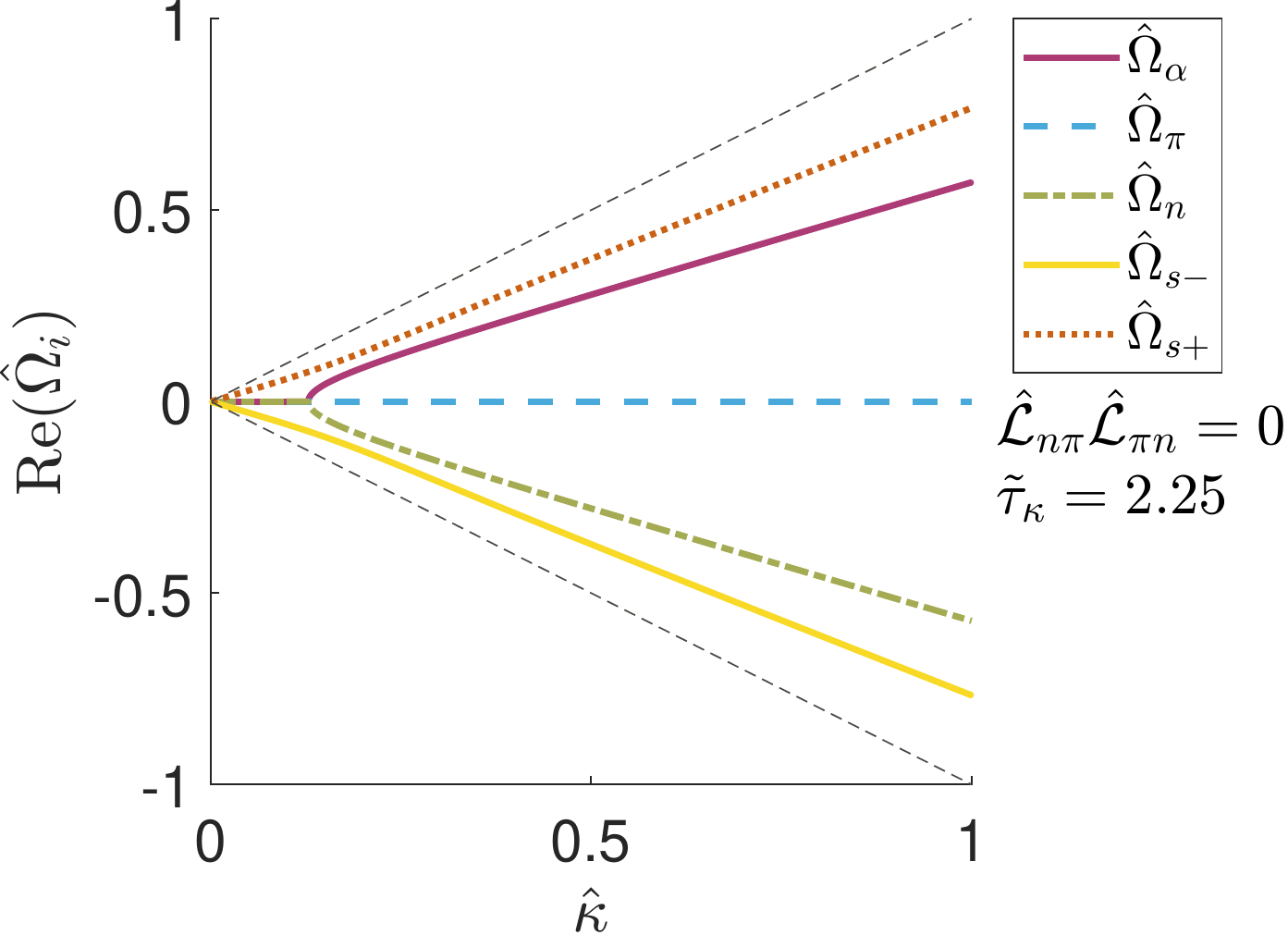}
	 \caption{The imaginary parts $\mathrm{Im}(\hat{\Omega}_i)$ (upper row) and 
	 the real parts $\mathrm{Re}(\hat{\Omega}_i)$ of the five longitudinal modes, $i \in \{s\pm, \pi, n, \alpha\}$, for
	 $\hat{\tau}_\pi = 5$, $\hat{\tau}_n = 27/4 = 6.75$, for the minimum and maximum values of 
	 $\tilde{\tau}_\kappa$, i.e., $9/16$ (left column)
	 and $9/4$ (right column). The sound modes $\hat{\Omega}_{s\pm}$ are shown by the solid yellow and
	 red dotted lines, the shear mode $\hat{\Omega}_\pi$ by the dashed blue, the charge-transport mode 
	 $\hat{\Omega}_\alpha$ by the solid magenta, and the charge-diffusion mode $\hat{\Omega}_n$ 
	 by the dash-dotted green line, respectively. In the lower row, the light cone is shown by the thin dashed
	 black lines.}
	 \label{figure1}
\end{figure*}

\subsection{Exploring the parameter space}

We are now ready to explore the space of parameters in the 
$\tilde{\tau}_\kappa - \hat{\mathcal{L}}_{n\pi} \hat{\mathcal{L}}_{\pi n}$ plane, see Fig.\ \ref{figure3}.
In this figure, one distinguishes several regions where the dispersion relations show different characteristics:
\begin{enumerate}
\item In the red region at the bottom, the asymptotic group velocity of the sound modes 
$\mathcal{T}_+ > 1$, i.e., the system is acausal. As discussed after Eq.\ (\ref{eq:83}) (cf.\ also Ref.\ \cite{Brito:2020nou}), 
while the system still appears stable in a background at rest, 
a moving background reveals that the system is actually unstable in this region. As discussed
above, the sound mode $\hat{\omega}_{+ -}(\hat{k})\equiv \hat{\omega}_{s-}(\hat{k})$
then becomes unstable.
\item In the gray region, condition (iii.1) is violated and, as shown above, the two sound modes 
become unstable. 
\item In the orange region, condition (iii.2) is violated, and the charge-diffusion and the charge-transport mode become
unstable. 
\item The dashed line is the upper boundary of the region where $\mathcal{M} \geq 0$.
Above this line, $\mathcal{T}_-$ becomes imaginary and the charge-transport and the charge-diffusion 
mode no longer propagate. They remain unstable modes. 
\end{enumerate}
Despite these various instabilities, the asymptotic group velocities for the propagating modes 
remain causal everywhere above the red region.

\begin{figure*}
	\includegraphics[scale=0.43]{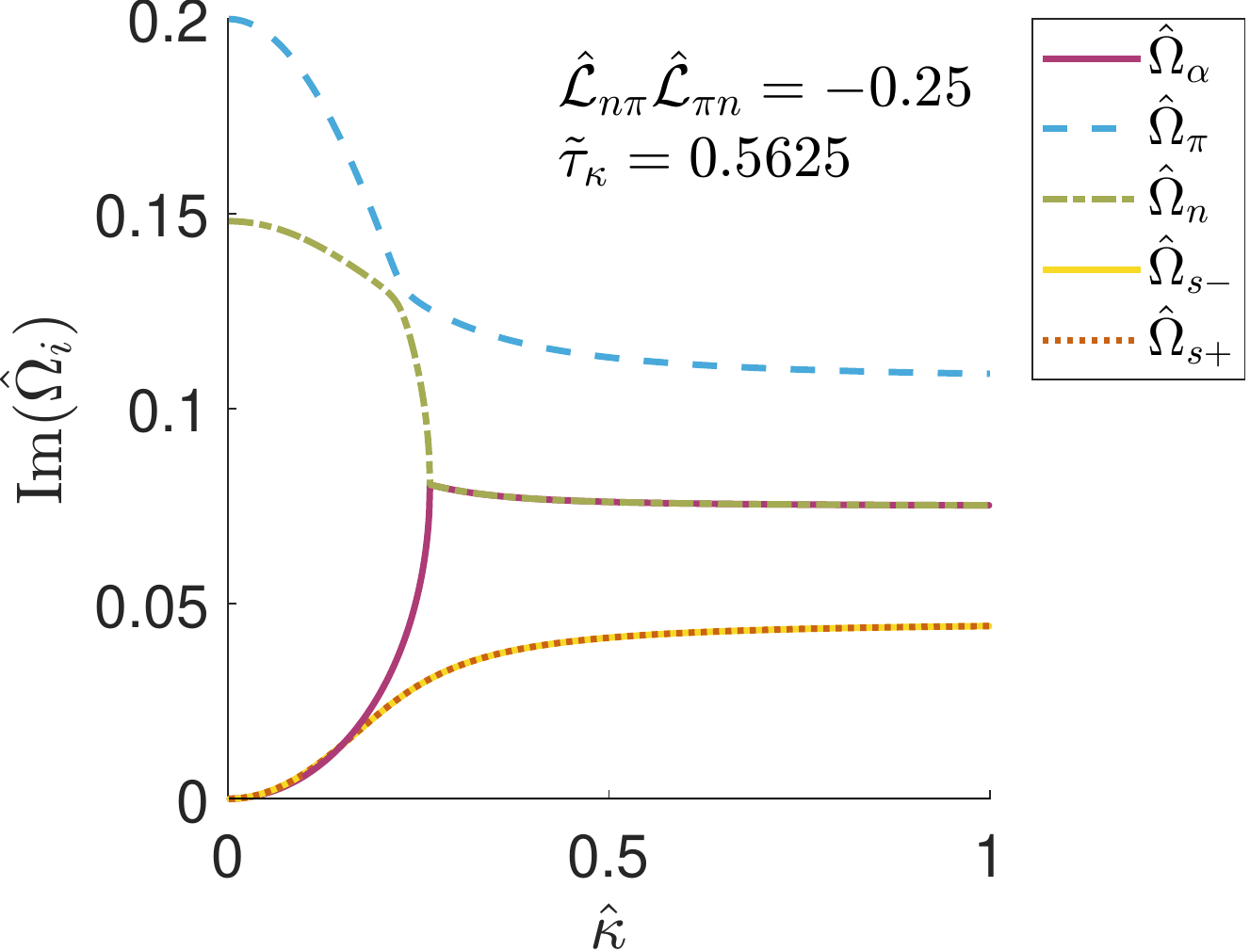}
	\includegraphics[scale=0.43]{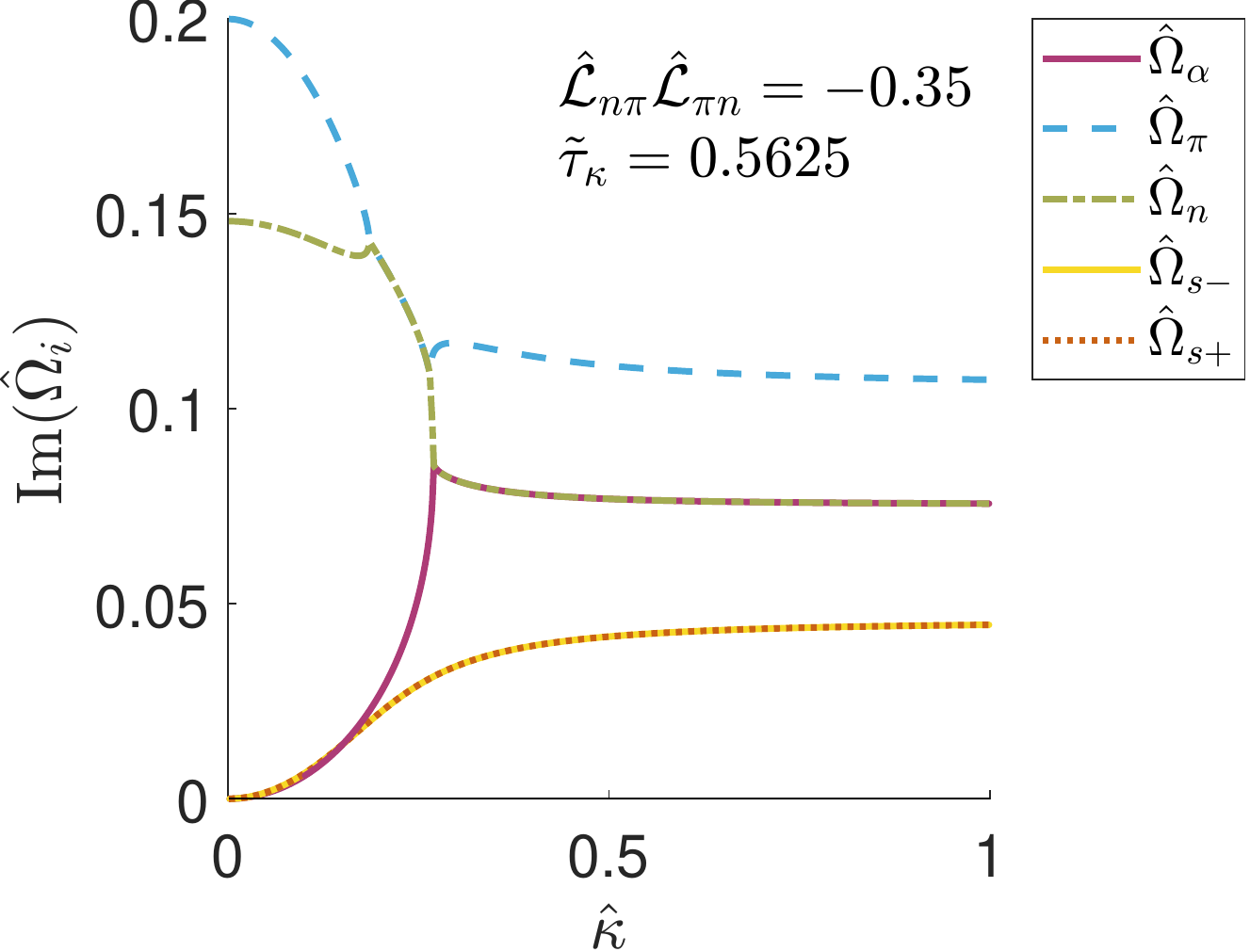}
	\includegraphics[scale=0.43]{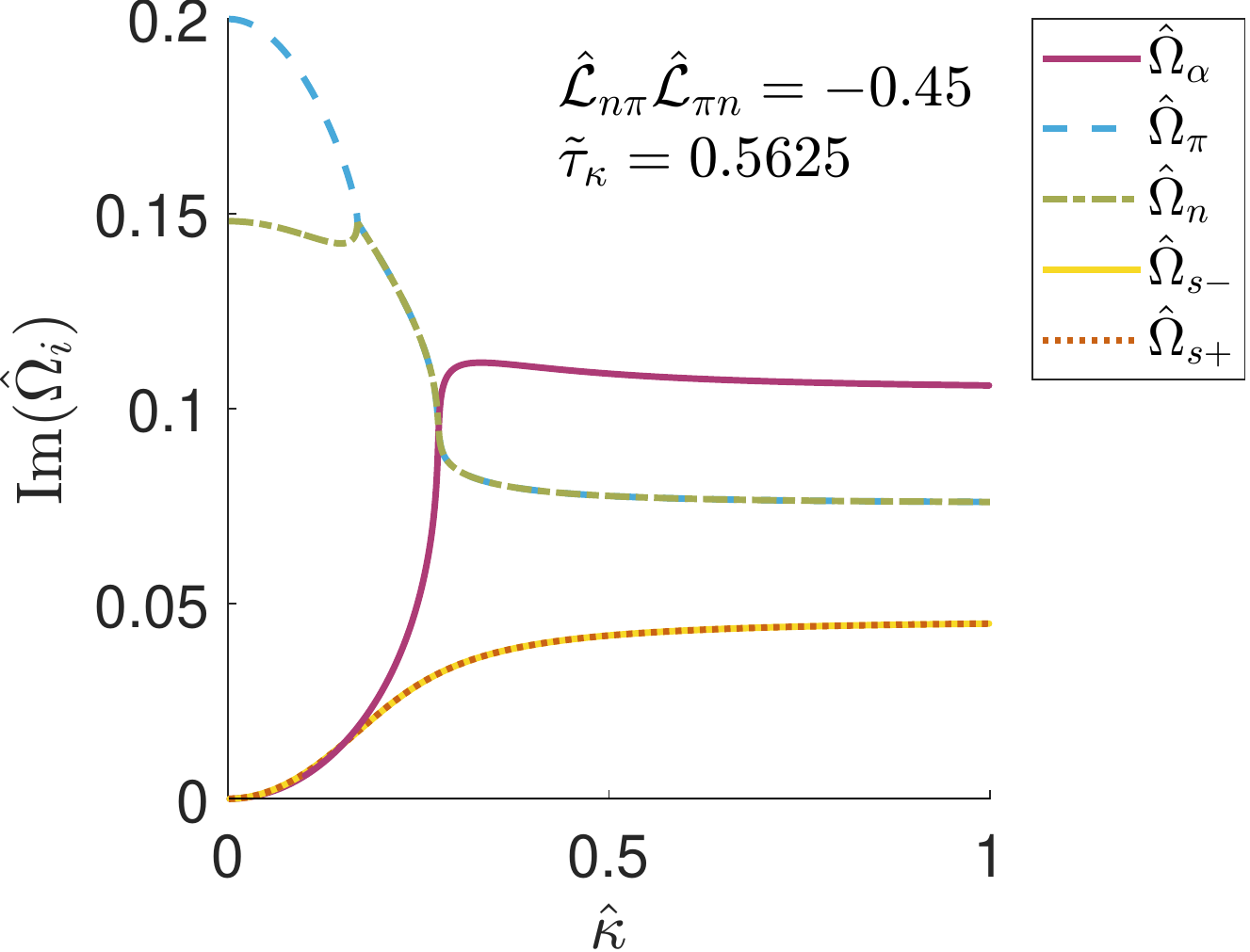}\\
	\includegraphics[scale=0.43]{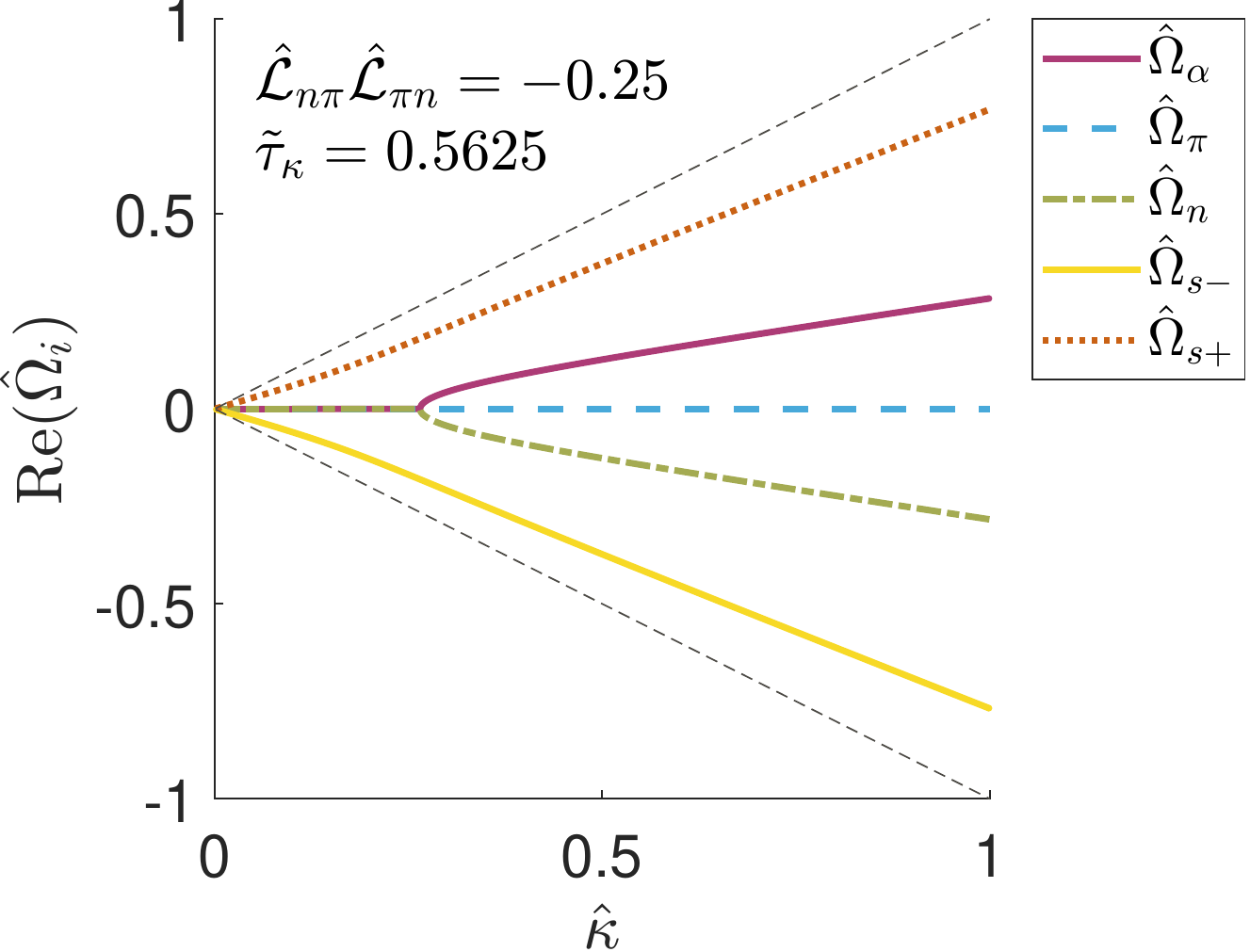}
	\includegraphics[scale=0.43]{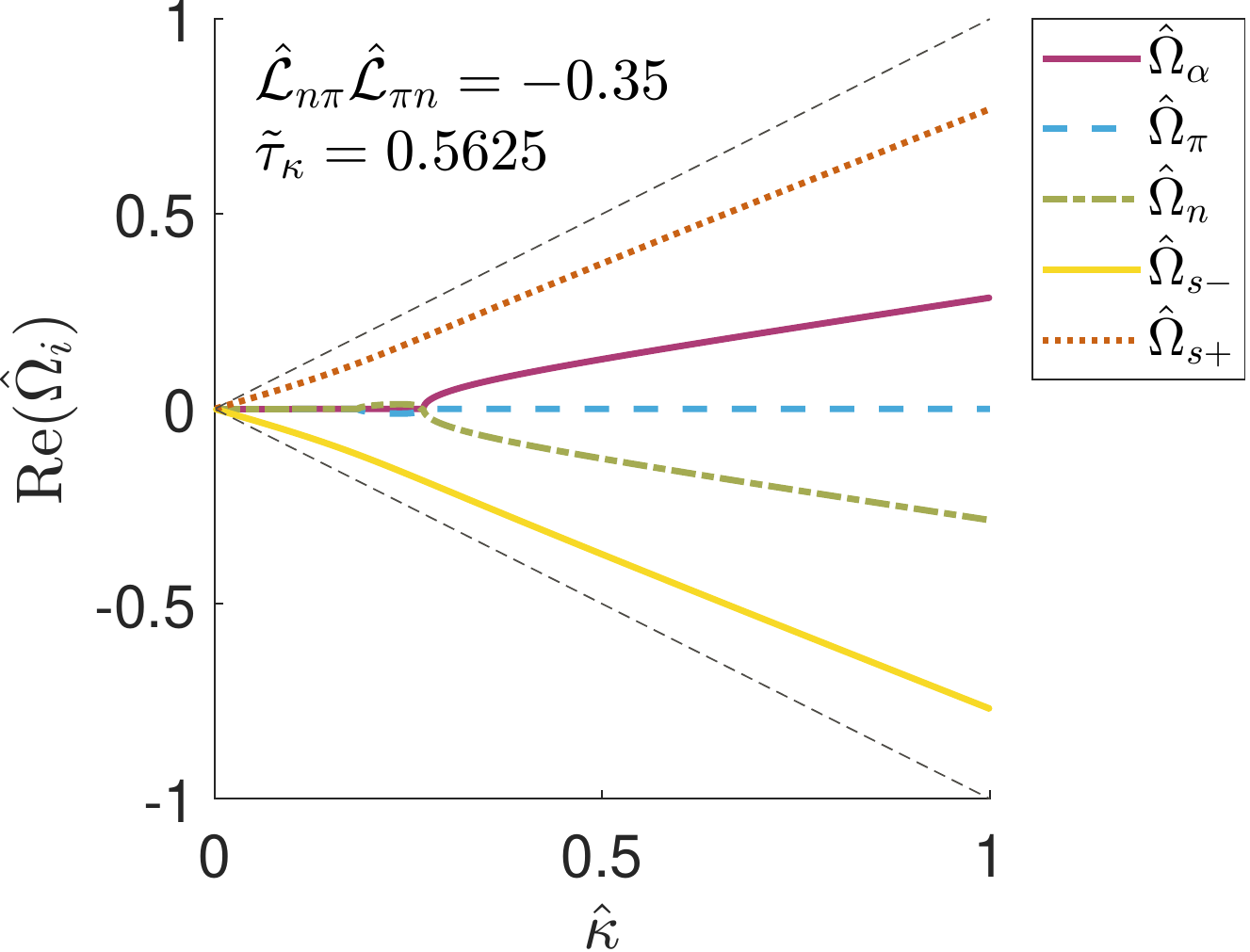}
	\includegraphics[scale=0.43]{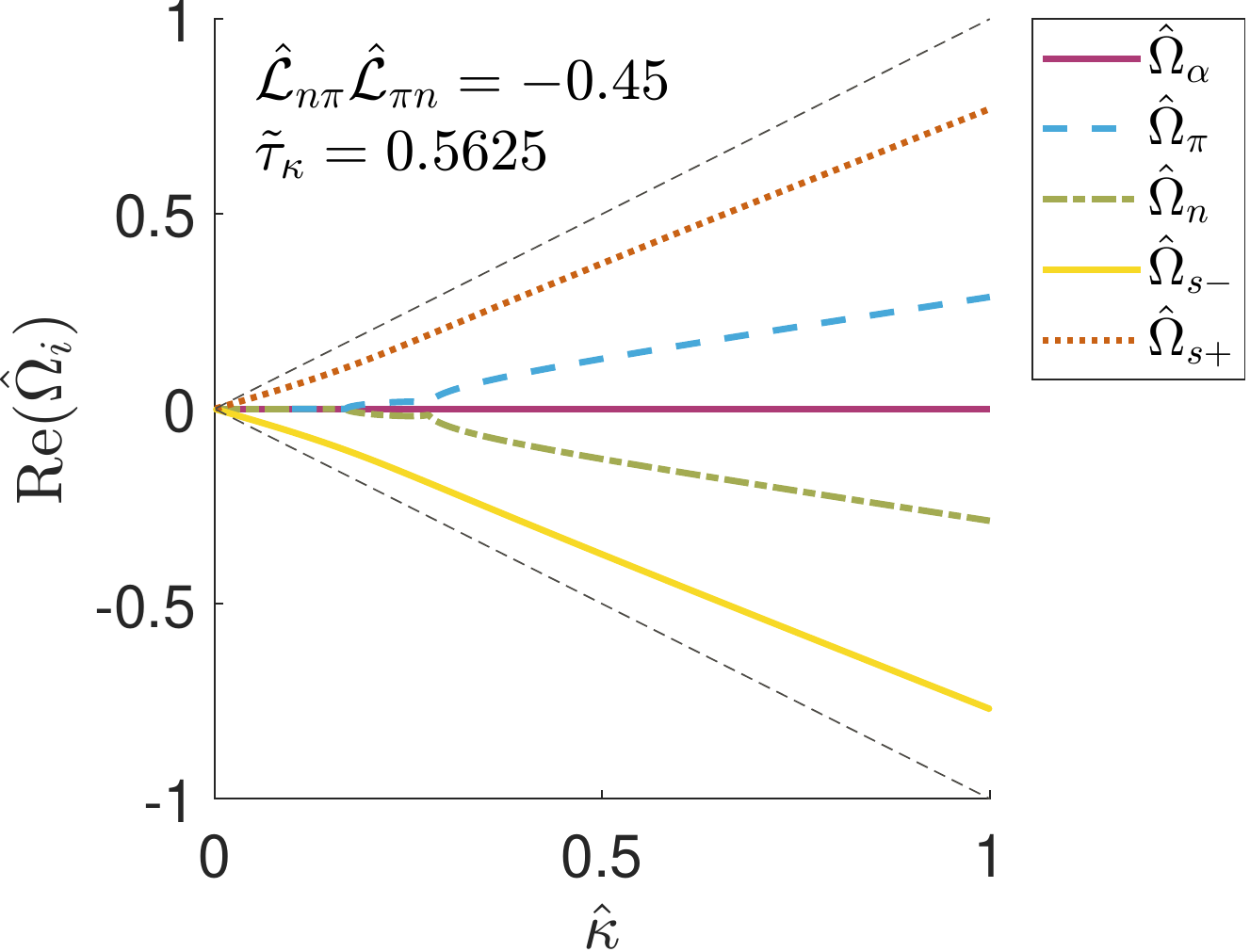}
	 \caption{The imaginary parts (upper row) and real parts (lower row) of the
	 five longitudinal modes, for $\hat{\tau}_\pi = 5$, $\hat{\tau}_n = 27/4 = 6.75$,  
	 the minimum value of $\tilde{\tau}_\kappa=9/16$, and the three values 
	 $\hat{\mathcal{L}}_{n\pi} \hat{\mathcal{L}}_{\pi n} =-0.25$ (left column), $-0.35$ (middle column), and
	 $-0.45$ (right column). Line modes are the same as in Fig.~\ref{figure1}.}
	 \label{figure2}
\end{figure*}

The system is stable and causal in the region above the line marked $\mathcal{T}_+ = 1$
and below the curves marked (iii.1) and (iii.2). This region can be further subdivided into three regions:
\begin{enumerate}
\item[5.] In the blue region, the dispersion relations behave as in the left column of Fig.\ \ref{figure2}, i.e.,
$\hat{\Omega}_n(\hat{\kappa})$ and $\hat{\Omega}_\alpha(\hat{\kappa})$ propagate above a certain critical value of 
$\hat{\kappa}$, while $\hat{\Omega}_\pi(\hat{\kappa})$ remains a nonpropagating
non-hydrodynamic mode. 
\item[6.] In the green region, the dispersion relations behave as in the right column of Fig.\ \ref{figure2},
i.e., $\hat{\Omega}_\pi(\hat{\kappa})$ and $\hat{\Omega}_n(\hat{\kappa})$ propagate, while $\hat{\Omega}_\alpha(\hat{\kappa})$ remains a nonpropagating
hydrodynamic mode. 
\item[7.] In the small white band between these two regions, the dispersion relations behave as in
the middle column of Fig.\ \ref{figure2}, i.e., there is a range of intermediate $\hat{\kappa}$ values where the 
imaginary parts of $\hat{\Omega}_\pi(\hat{\kappa})$ and $\hat{\Omega}_n(\hat{\kappa})$ become degenerate, while
for larger values of $\hat{\kappa}$ the imaginary parts of $\hat{\Omega}_n(\hat{\kappa})$ and $\hat{\Omega}_\alpha(\hat{\kappa})$ 
become degenerate. Whenever this happens, these modes develop a nonzero real part and start to propagate.
\end{enumerate}
That the system behaves differently in these three regions can only be found by an explicit calculation of the
dispersion relations. We have already summarized our findings here; a more detailed discussion follows below.

In the following subsections, we will demonstrate the validity of these results by explicitly calculating the
dispersion relations at selected points in the $\tilde{\tau}_\kappa - \hat{\mathcal{L}}_{n\pi} \hat{\mathcal{L}}_{\pi n}$ plane. This is done by numerically solving Eq.\ (\ref{eq:60}) with a root-finding algorithm.
As was shown above, a nonvanishing background velocity does not influence those boundaries in this plane which can be calculated 
analytically, i.e., the solid red, orange, and black curves as well as the dotted line. An exception is the white band inside the stable region. Our numerical solution of Eq.\ (\ref{eq:60}) shows that, for $V>0$, this band tends to curve upwards and
gets wider as $\tilde{\tau}_\kappa$ is increased. However, we do not discuss this in further detail and restrict the following discussion to the case $V=0$. 
We note that our results complement the investigations
of Ref.~\cite{Brito:2020nou}, where in particular the violation of conditions (iii.1) and (iii.2) was not further investigated.

\subsubsection{$\hat{\mathcal{L}}_{n\pi} \hat{\mathcal{L}}_{\pi n} =0$}

In order to see the effect of varying
$\tilde{\tau}_\kappa$ from 9/16 to 9/4 most clearly, we first set $\hat{\mathcal{L}}_{n\pi} \hat{\mathcal{L}}_{\pi n} =0$. 
In this case, the last term in Eq.~(\ref{long_modes_2}) vanishes and the sound and shear modes result from the roots of
the third-order polynomial in the first bracket, while the charge-transport and diffusion modes are given by the 
roots of the second-order polynomial in the second bracket.
All roots can be given in closed analytical form~\cite{Ambrus:2017keg}, but we refrain from quoting them
explicity for the sake of brevity.

The real and imaginary parts of the five longitudinal modes $\hat{\Omega}_{s\pm}(\hat{\kappa})$, $\hat{\Omega}_\pi(\hat{\kappa})$, 
$\hat{\Omega}_n(\hat{\kappa})$, and $\hat{\Omega}_\alpha(\hat{\kappa})$ are shown as a function of  
$\hat{\kappa}$ in Fig.~\ref{figure1}. The effect of varying the background charge $n_0$, 
i.e., varying the coefficient $\tilde{\tau}_\kappa$, can be
seen by comparing the left and the right columns. 
We only show results for the minimum value $\tilde{\tau}_\kappa = 9/16$
and the maximum value $\tilde{\tau}_\kappa = 9/4$. The first case agrees with the results shown in Fig.~4 of
Ref.~\cite{Brito:2020nou}. One observes that a larger value of $\tilde{\tau}_\kappa$ simply reduces the
value of $\hat{\kappa}$ where the imaginary parts of the charge-transport and charge-diffusion modes
become degenerate and these two modes start to propagate (as indicated by a nonvanishing real part). 
For the case $\hat{\mathcal{L}}_{n\pi} \hat{\mathcal{L}}_{\pi n} =0$, 
the sound and shear modes are, of course, not affected by a variation of $\tilde{\tau}_\kappa $.

The sound modes are propagating modes for any value of $\hat{\kappa}$, which simply become attenuated
at finite values of $\hat{\kappa}$. On the other hand, the charge-transport and
charge-diffusion modes become propagating only for sufficiently large values of $\hat{\kappa}$. In contrast,
the shear mode never propagates and is purely diffusive. All modes are stable, as they all feature manifestly
non-negative imaginary parts.

\begin{figure}
	\includegraphics[scale=0.5]{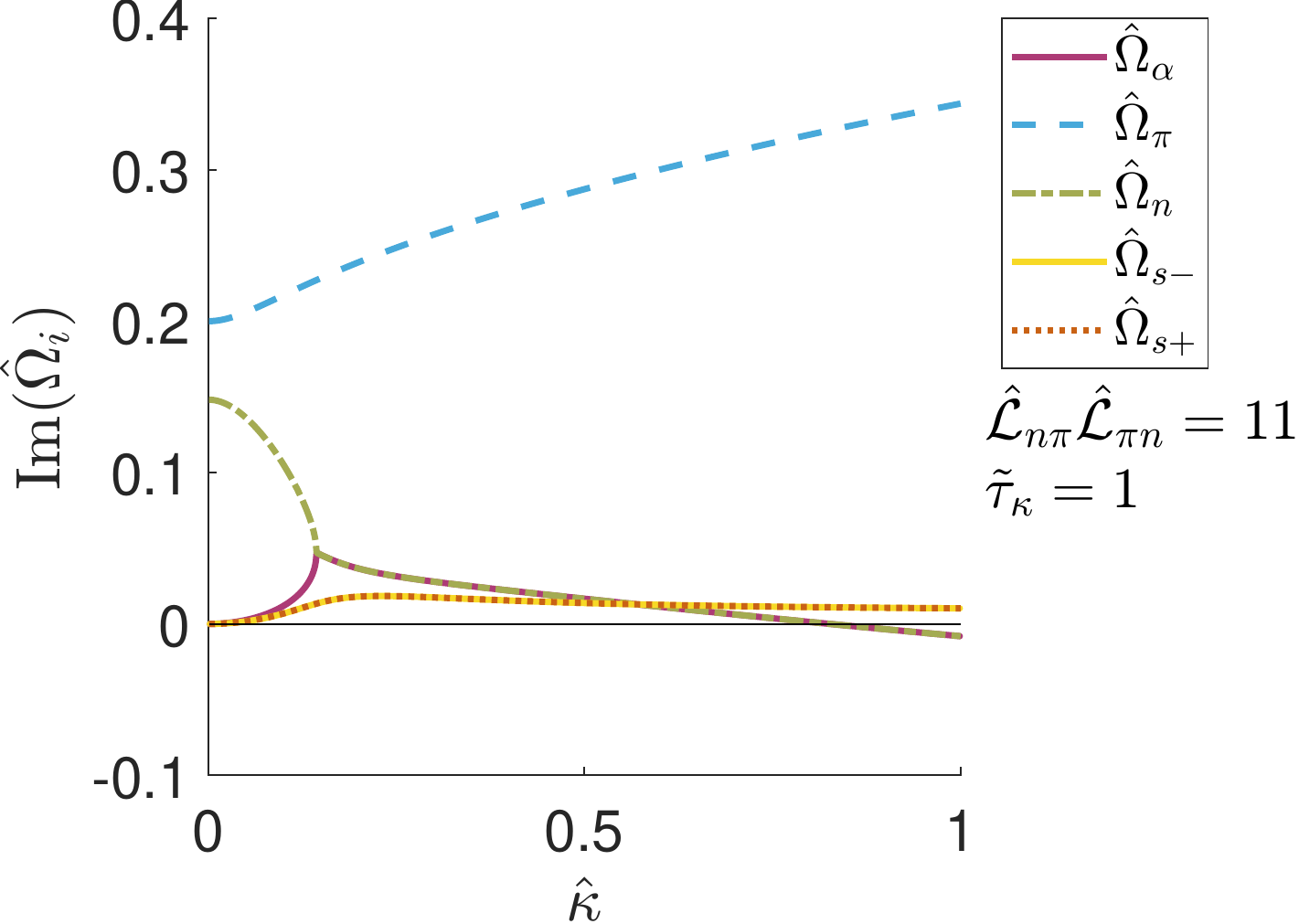}
	\includegraphics[scale=0.5]{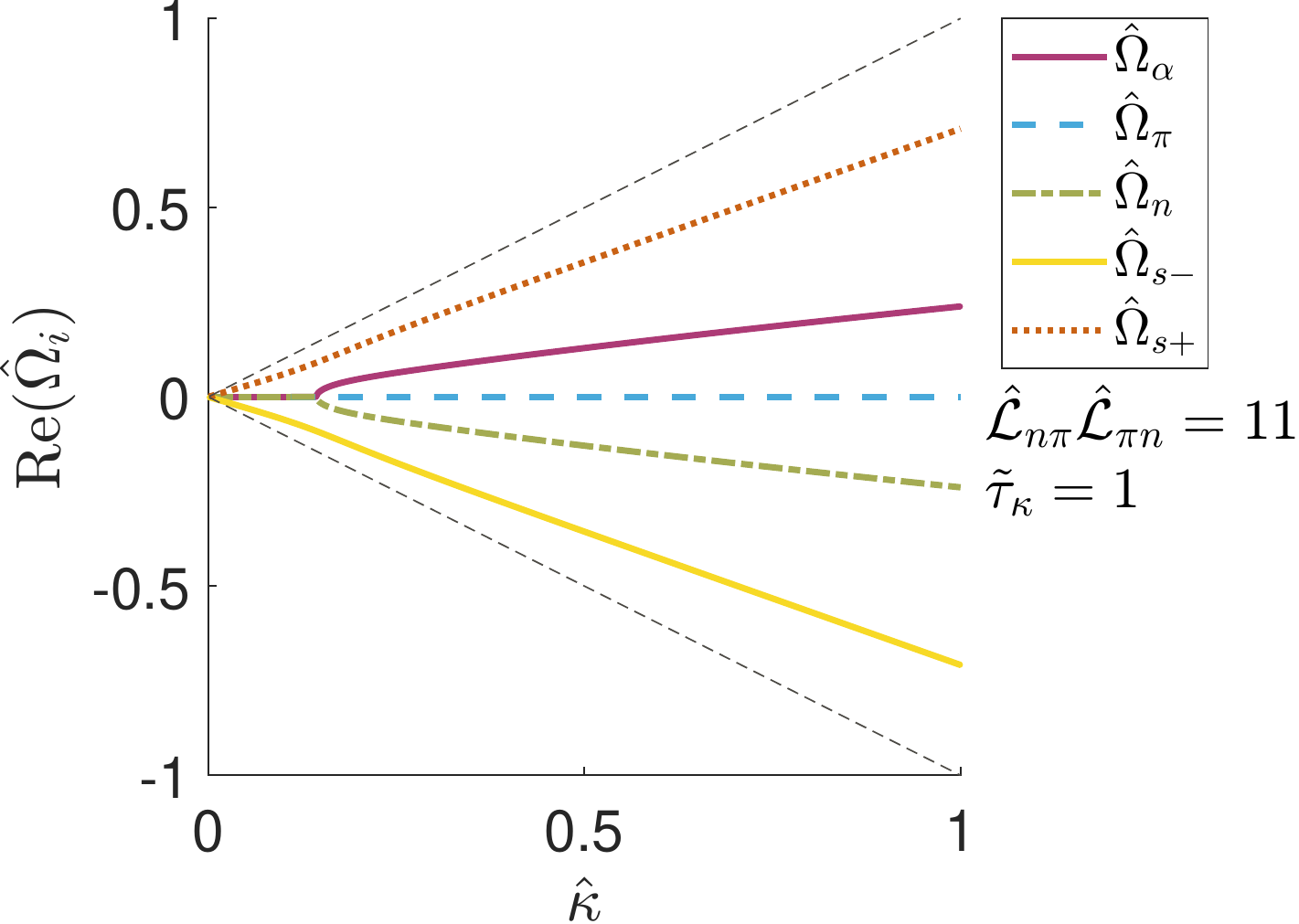}
	 \caption{The imaginary parts (upper panel) and real parts (lower panel) of the
	 five longitudinal modes, for $\hat{\tau}_\pi = 5$, $\hat{\tau}_n = 27/4 = 6.75$,  
	 $\tilde{\tau}_\kappa=1$, and 
	 $\hat{\mathcal{L}}_{n\pi} \hat{\mathcal{L}}_{\pi n} =11$ 
  (corresponding to point (a) in Fig.\ \ref{figure3}). Line modes are the same as in Fig.~\ref{figure1}.}
	 \label{figure4}
\end{figure}

\begin{figure}
	\hspace*{-0.6cm} \includegraphics[scale=0.5]{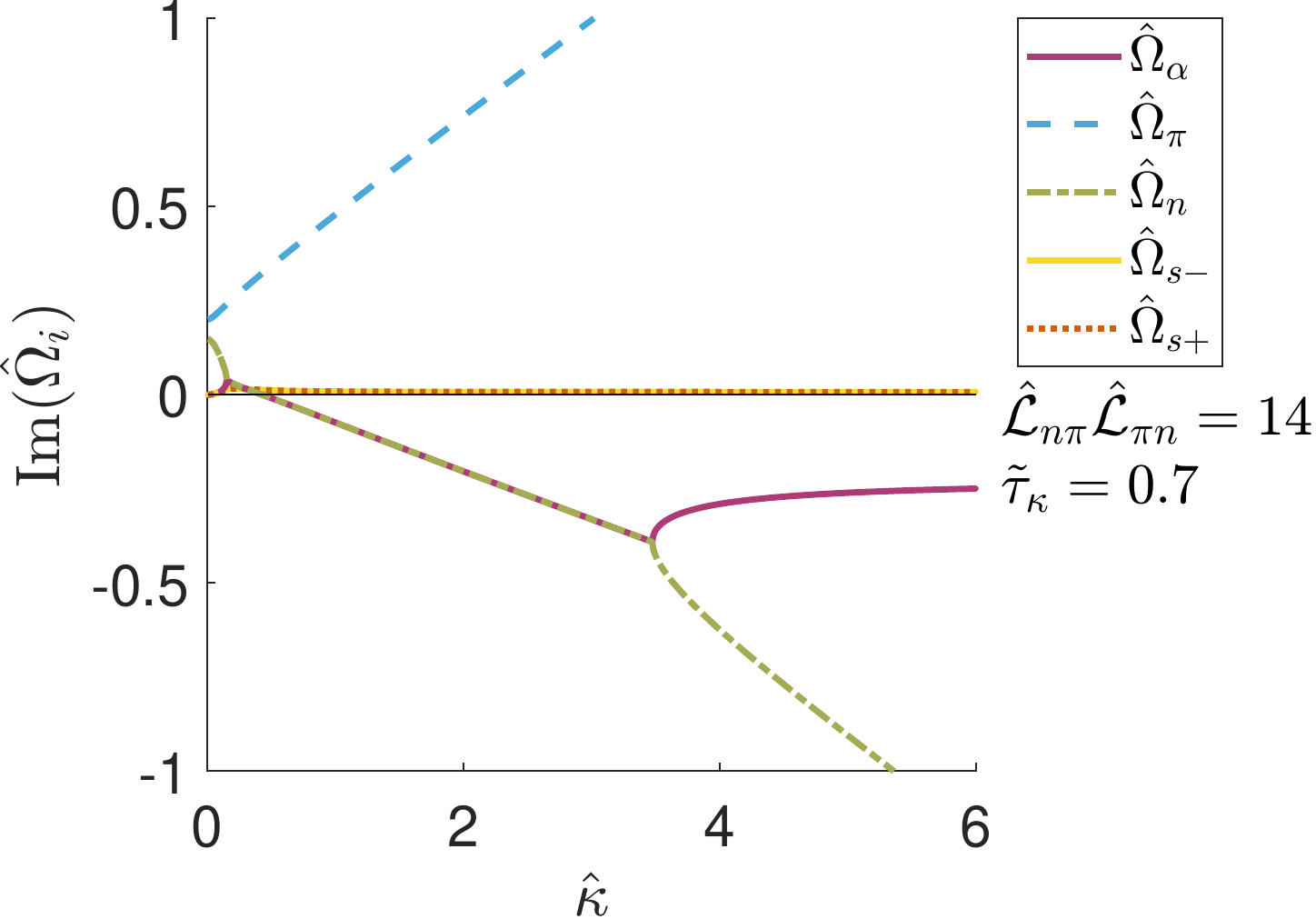}
	\includegraphics[scale=0.5]{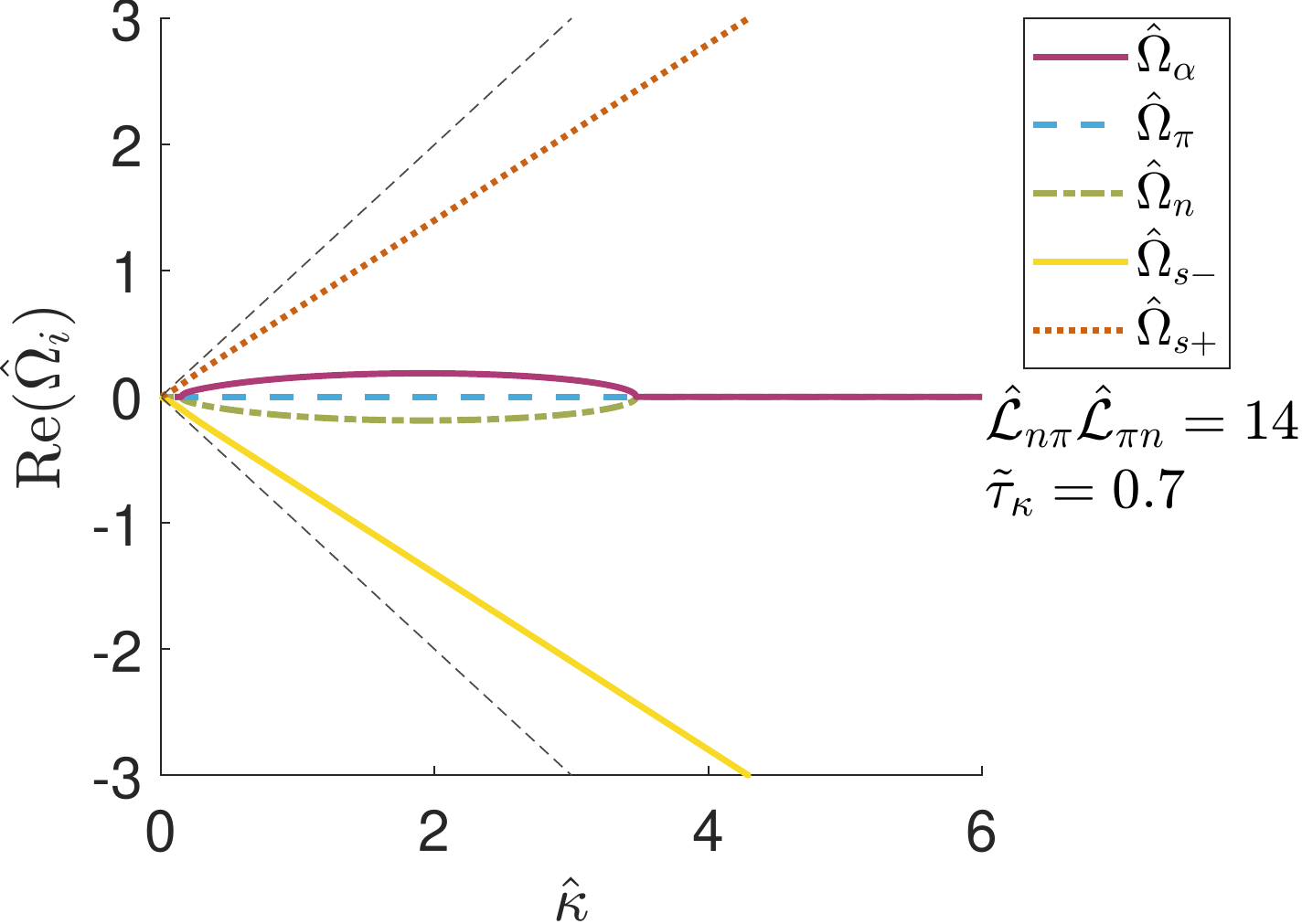}
	 \caption{The imaginary parts (upper panel) and real parts (lower panel) of the
	 five longitudinal modes, for $\hat{\tau}_\pi = 5$, $\hat{\tau}_n = 27/4 = 6.75$,  
	 $\tilde{\tau}_\kappa=0.7$, and 
	 $\hat{\mathcal{L}}_{n\pi} \hat{\mathcal{L}}_{\pi n} =14$ 
  (corresponding to point (b) in Fig.\ \ref{figure3}). Line modes are the same as in Fig.~\ref{figure1}.}
	 \label{figure5}
\end{figure}

\subsubsection{The stable region for $\hat{\mathcal{L}}_{n\pi} \hat{\mathcal{L}}_{\pi n} \neq 0$}

We now investigate the stable region for nonzero values of
$\hat{\mathcal{L}}_{n\pi} \hat{\mathcal{L}}_{\pi n}$. To this end, we first
keep $\tilde{\tau}_{\kappa}$ at its minimum value of 9/16 and take  
$\hat{\mathcal{L}}_{n\pi} \hat{\mathcal{L}}_{\pi n}$ to be $-0.25$, $-0.35$, and $-0.45$. The first value
lies in the blue, the second in the white, and the third in the green region of Fig.\ \ref{figure3}.
The results are shown in Fig.~\ref{figure2}.

Comparing the left column of Fig.~\ref{figure2} with that of Fig.~\ref{figure1}, we first observe that decreasing
the coupling term $\hat{\mathcal{L}}_{n\pi} \hat{\mathcal{L}}_{\pi n}$ from zero to $-0.25$ leads 
to an attraction of the imaginary parts of the shear and the charge-diffusion modes. Further decreasing 
$\hat{\mathcal{L}}_{n\pi} \hat{\mathcal{L}}_{\pi n}$ to $-0.35$ we find that the imaginary parts of these two modes
become degenerate for an intermediate range of $\hat{\kappa}$ values, see middle column of Fig.~\ref{figure2}. 
In this range, they assume small, but nonvanishing real parts, cf.\ lower middle panel of Fig.~\ref{figure2},
indicating that they become propagating modes. Further increasing $\hat{\kappa}$, the degeneracy
of the imaginary parts of shear and charge-diffusion modes is lifted. The shear mode remains
nonpropagating for all larger values of $\hat{\kappa}$, while the imaginary parts of
the charge-diffusion and charge-transport modes again become degenerate, just as in the previously discussed cases.
Once this degeneracy lifted, these modes start to propagate.

Further decreasing $\hat{\mathcal{L}}_{n\pi} \hat{\mathcal{L}}_{\pi n}$ to $-0.45$, something peculiar happens:
as shown in the right column of Fig.~\ref{figure2}, the imaginary parts of the shear and the charge-diffusion modes 
again become degenerate at some value of $\hat{\kappa}$. However, this degeneracy is never lifted
when increasing $\hat{\kappa}$, and these modes remain propagating modes for all larger
values of $\hat{\kappa}$. In this case the charge-transport mode is the one which remains purely diffusive,
instead of the shear mode which was purely diffusive for vanishing $\hat{\mathcal{L}}_{n\pi} \hat{\mathcal{L}}_{\pi n}$.
We remark that this behavior has also been seen
in Fig.~13 of Ref.~\cite{Brito:2020nou} (for $\hat{\mathcal{L}}_{n\pi} \hat{\mathcal{L}}_{\pi n} = -1$). The transition
to this behavior, however, was not investigated by systematically decreasing 
$\hat{\mathcal{L}}_{n\pi} \hat{\mathcal{L}}_{\pi n}$ from zero to larger negative values, and our results can be seen
as completing the discussion.
This peculiar behavior can be observed also for larger values of $\tilde{\tau}_\kappa$, but then the
values of $\hat{\mathcal{L}}_{n\pi} \hat{\mathcal{L}}_{\pi n}$ where this happens (i.e., the white region in  
Fig.\ \ref{figure3}) decrease.

\subsubsection{The unstable region for large $\hat{\mathcal{L}}_{n\pi} \hat{\mathcal{L}}_{\pi n}$}

In the unstable region for large $\hat{\mathcal{L}}_{n\pi} \hat{\mathcal{L}}_{\pi n}$ we identify five regions where
the dispersion relations behave differently. We select representative values for $\tilde{\tau}_\kappa$
and $\hat{\mathcal{L}}_{n\pi} \hat{\mathcal{L}}_{\pi n}$ in these regions, shown by the crosses
marked (a) -- (e) in Fig.\ \ref{figure3}, and explicitly compute the 
dispersion relations at these points.

\textit{Point (a)} is located at $(\tilde{\tau}_\kappa, \hat{\mathcal{L}}_{n\pi} \hat{\mathcal{L}}_{\pi n}) = (1,11)$ and the
corresponding dispersion relations are shown in Fig.\ \ref{figure4}. As expected from the discussion of Fig.\ \ref{figure3}, at this point only the charge-transport and charge-diffusion modes become unstable as $\hat{\kappa}$ increases. For the chosen parameters this happens around $\hat{\kappa} \simeq 0.8$. The other modes all remain stable. In particular, the shear mode 
$\hat{\Omega}_\pi (\hat{\kappa})$ is identified with the nonpropagating mode $\hat{\Omega}_0(\hat{\kappa})$ from
the analysis of Eq.\ (\ref{eq:60}) at
asymptotically large values of $\hat{\kappa}$.

\textit{Point (b)} is located at $(\tilde{\tau}_\kappa, \hat{\mathcal{L}}_{n\pi} \hat{\mathcal{L}}_{\pi n}) = (0.7,14)$ and the
corresponding dispersion relations are shown in Fig.\ \ref{figure5}. As we are in the region above the dotted line
in Fig.\ \ref{figure3}, i.e., where
$\mathcal{T}_-$ becomes imaginary,
we expect that, for $\hat{\kappa} \to \infty$, apart from the nonpropagating
mode $\hat{\Omega}_0 (\hat{\kappa})$
we have two nonpropagating modes
$\hat{\Omega}_{-\pm} (\hat{\kappa})$ with imaginary parts $\sim \pm \hat{\kappa}$. Figure \ref{figure5} confirms this expectation, but contains further details at intermediate values of $\hat{\kappa}$ which did not emerge in the asymptotic analysis of 
Eq.\ (\ref{eq:60}). The shear mode is always a nonpropagating mode and its imaginary part indeed grows $\sim \hat{\kappa}$ for large $\hat{\kappa}$. Therefore, asymptotically $\hat{\Omega}_\pi (\hat{\kappa}) \equiv \hat{\Omega}_{-+} (\hat{\kappa})$. 
The two unstable modes are again the charge-transport and charge-diffusion modes, and for $\hat{\kappa} \to \infty$ we have $\hat{\Omega}_\alpha (\hat{\kappa}) \equiv \hat{\Omega}_0 (\hat{\kappa})$
and $\hat{\Omega}_n (\hat{\kappa}) \equiv \hat{\Omega}_{--} (\hat{\kappa})$, with
both modes nonpropagating.
However, there is a region of intermediate values of $\hat{\kappa}$
where the imaginary parts of the latter modes become degenerate and they develop nonvanishing real parts, i.e., are propagating modes.

\begin{figure}
	\hspace*{-0.6cm} \includegraphics[scale=0.5]{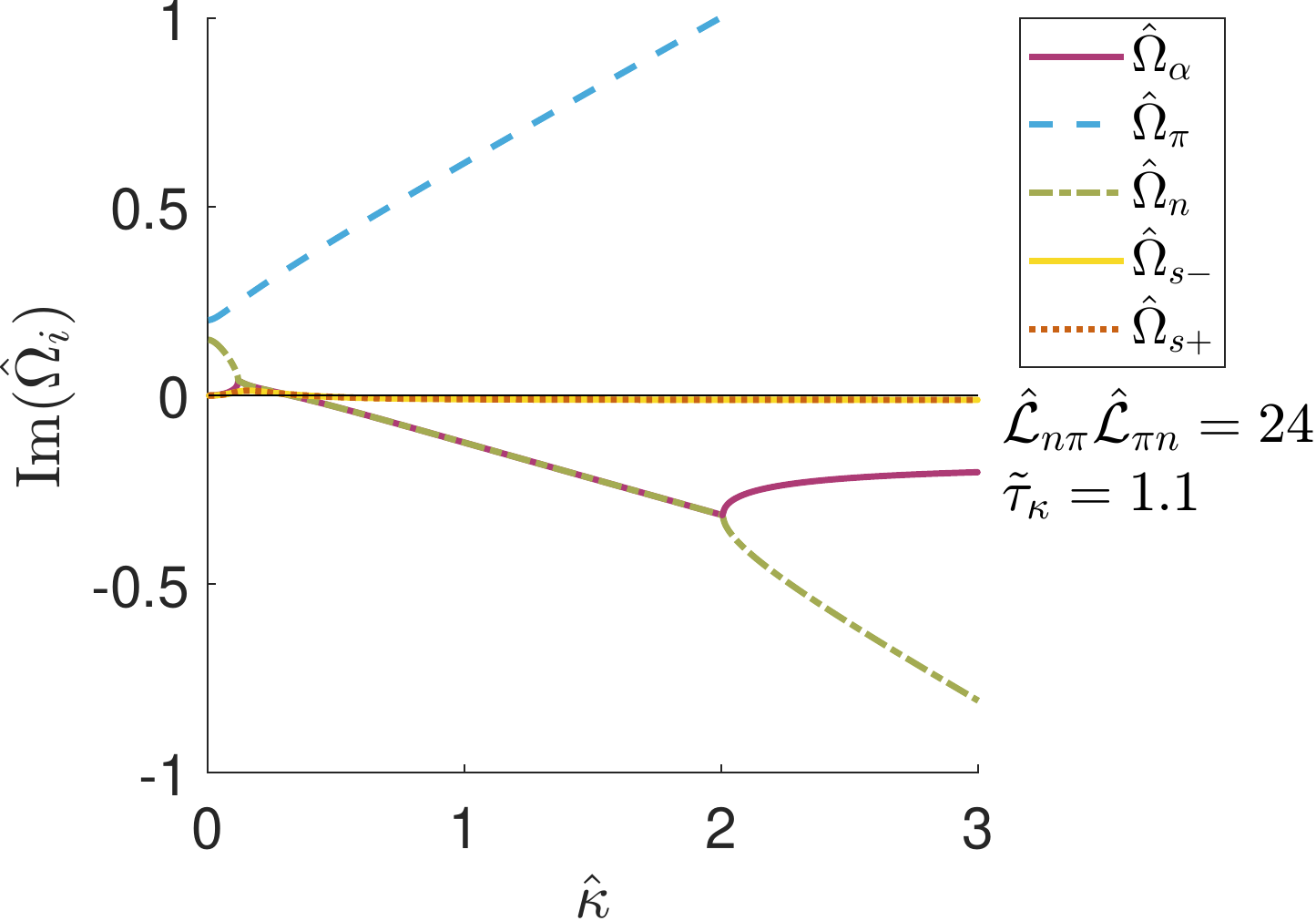}
	\includegraphics[scale=0.5]{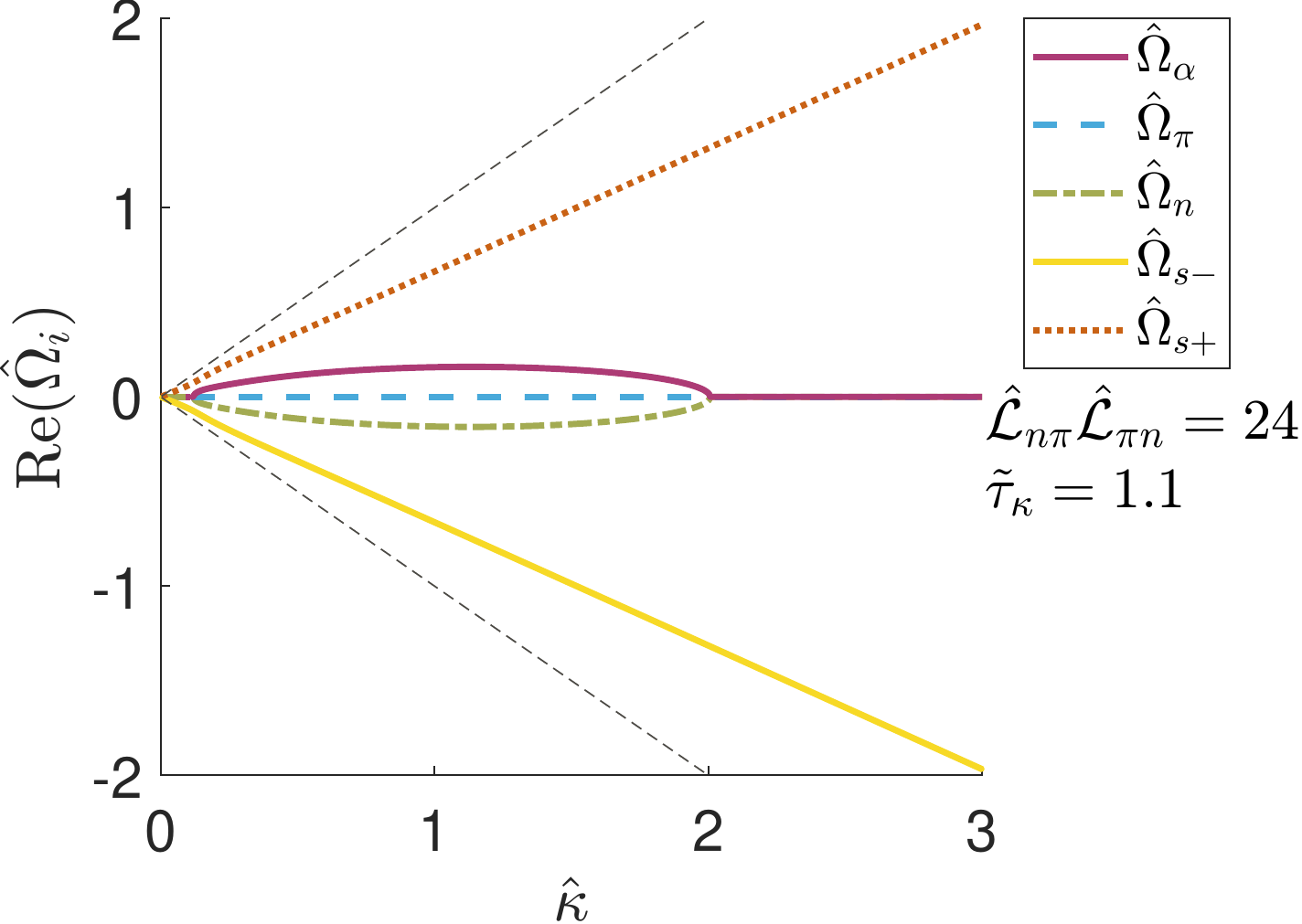}
	 \caption{The imaginary parts (upper panel) and real parts (lower panel) of the
	 five longitudinal modes, for $\hat{\tau}_\pi = 5$, $\hat{\tau}_n = 27/4 = 6.75$,  
	 $\tilde{\tau}_\kappa=1.1$, and 
	 $\hat{\mathcal{L}}_{n\pi} \hat{\mathcal{L}}_{\pi n} =24$ 
  (corresponding to point (c) in Fig.\ \ref{figure3}). Line modes are the same as in Fig.~\ref{figure1}.}
	 \label{figure6}
\end{figure}

\textit{Point (c)} is located at $(\tilde{\tau}_\kappa, \hat{\mathcal{L}}_{n\pi} \hat{\mathcal{L}}_{\pi n}) = (1.1,24)$ and the
corresponding dispersion relations are shown in Fig.\ \ref{figure6}. These closely resemble the situation in Fig.\ \ref{figure5}, however, with the major difference that now also the two sound modes become unstable, since we are in the region in Fig.\ \ref{figure3} where condition (iii.1) is violated. Comparing Figs.\ \ref{figure5} and \ref{figure6} one notices that in the former figure, the imaginary parts of these modes are slightly above the zero line, and thus they are stable, while in the latter figure they are slightly below, and thus unstable. This confirms all expectations from our analysis in Sec.\ \ref{sec:III}.

\begin{figure}
	\includegraphics[scale=0.5]{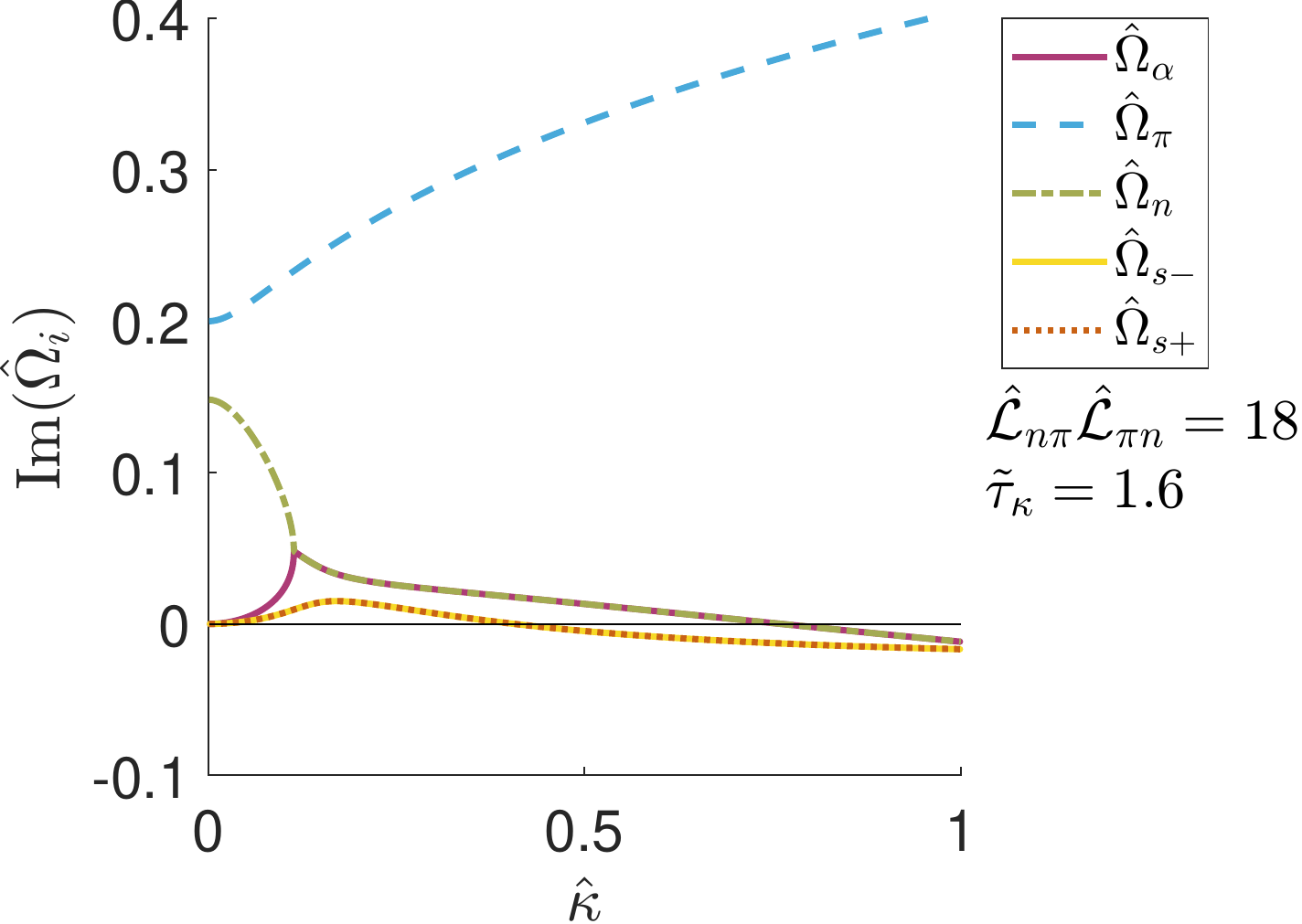}
	\includegraphics[scale=0.5]{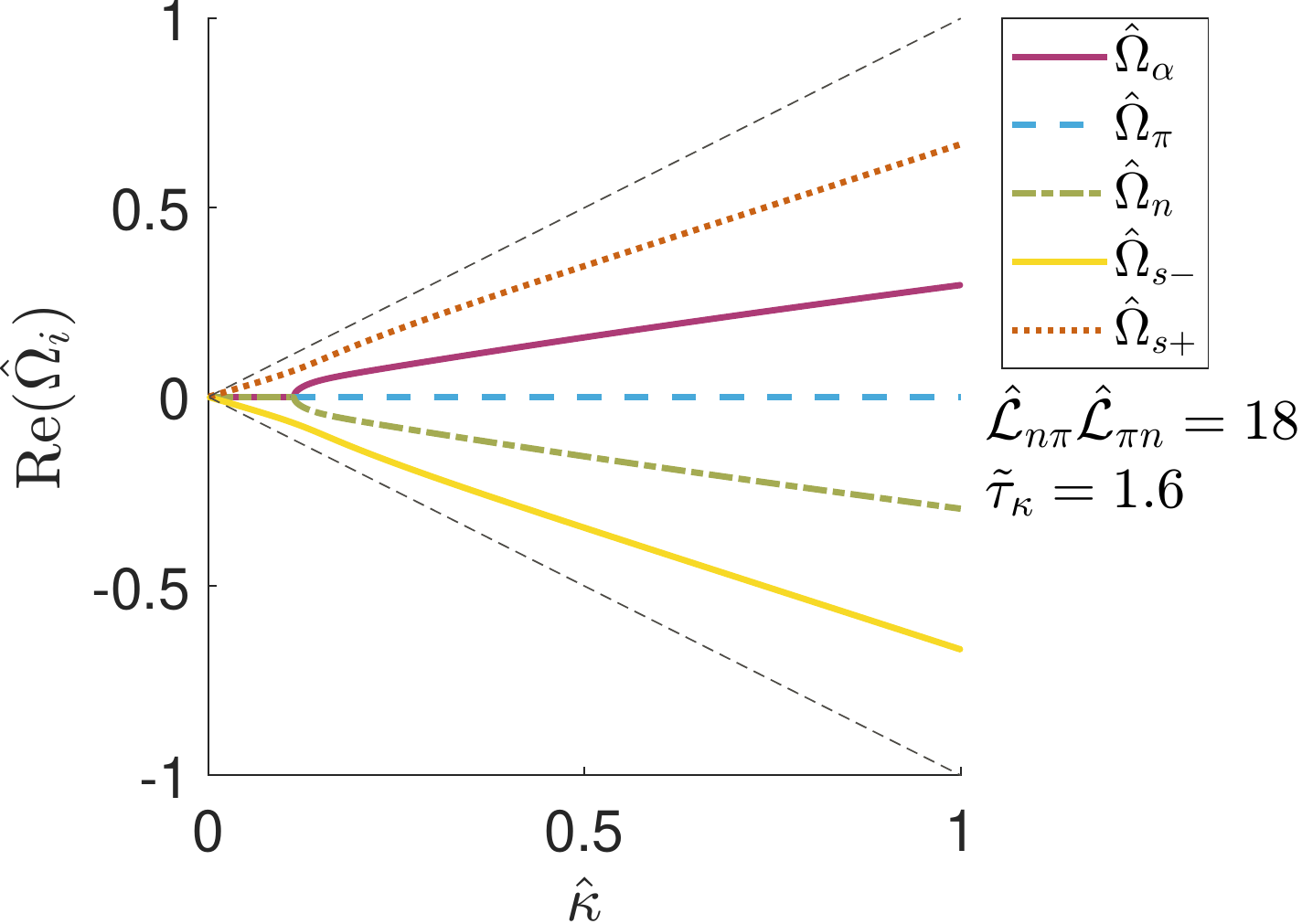}
	 \caption{The imaginary parts (upper panel) and real parts (lower panel) of the
	 five longitudinal modes, for $\hat{\tau}_\pi = 5$, $\hat{\tau}_n = 27/4 = 6.75$,  
	 $\tilde{\tau}_\kappa=1.6$, and 
	 $\hat{\mathcal{L}}_{n\pi} \hat{\mathcal{L}}_{\pi n} =18$ 
  (corresponding to point (d) in Fig.\ \ref{figure3}). Line modes are the same as in Fig.~\ref{figure1}.}
	 \label{figure7}
\end{figure}

\textit{Point (d)} is located at $(\tilde{\tau}_\kappa, \hat{\mathcal{L}}_{n\pi} \hat{\mathcal{L}}_{\pi n}) = (1.6,18)$ and the
corresponding dispersion relations are shown in Fig.\ \ref{figure7}. Here, the shear mode is the stable, 
nonpropagating mode
$\hat{\Omega}_0 (\hat{\kappa})$, while all other modes are propagating, but unstable modes.

\begin{figure}
	\includegraphics[scale=0.5]{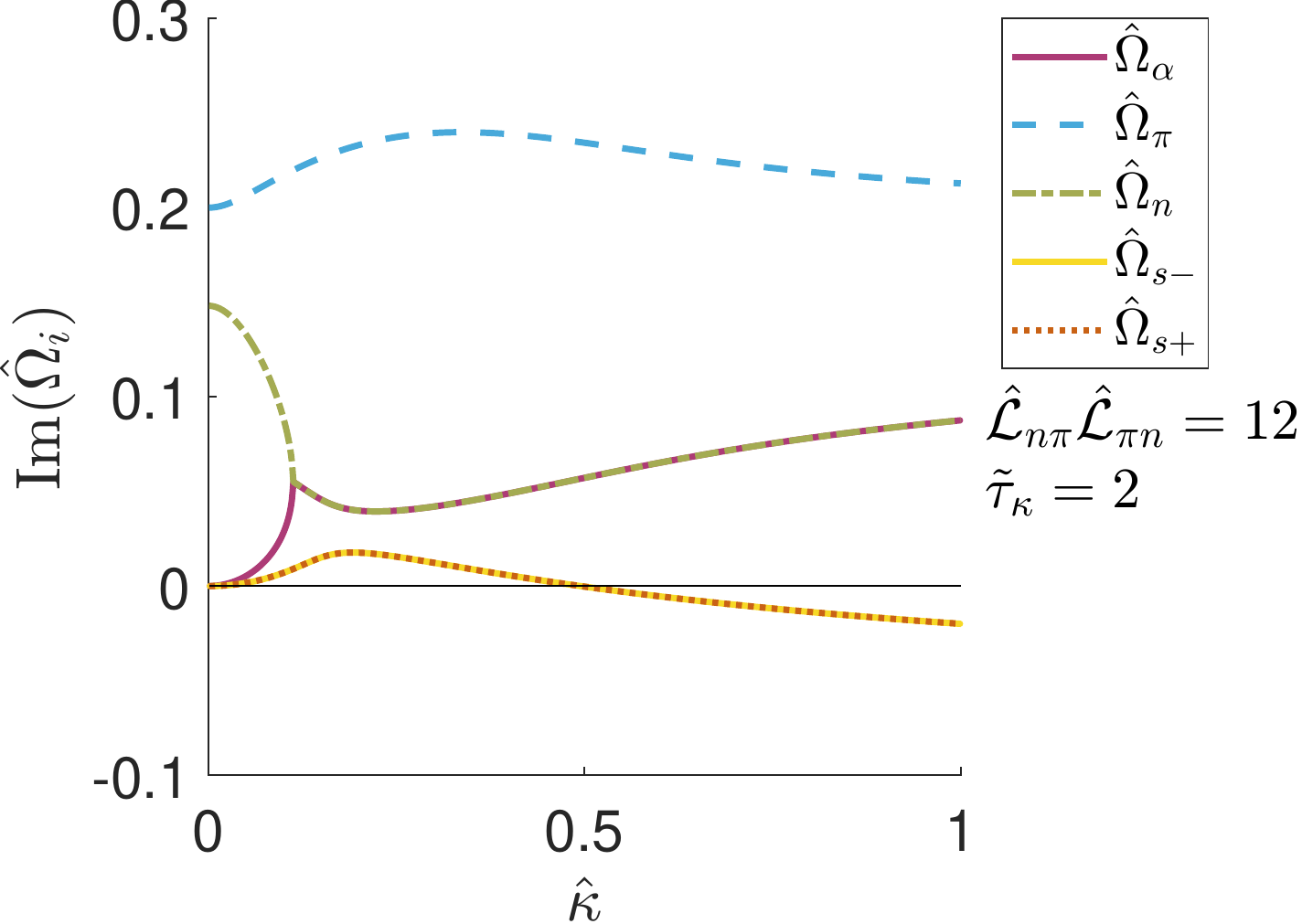}
	\includegraphics[scale=0.5]{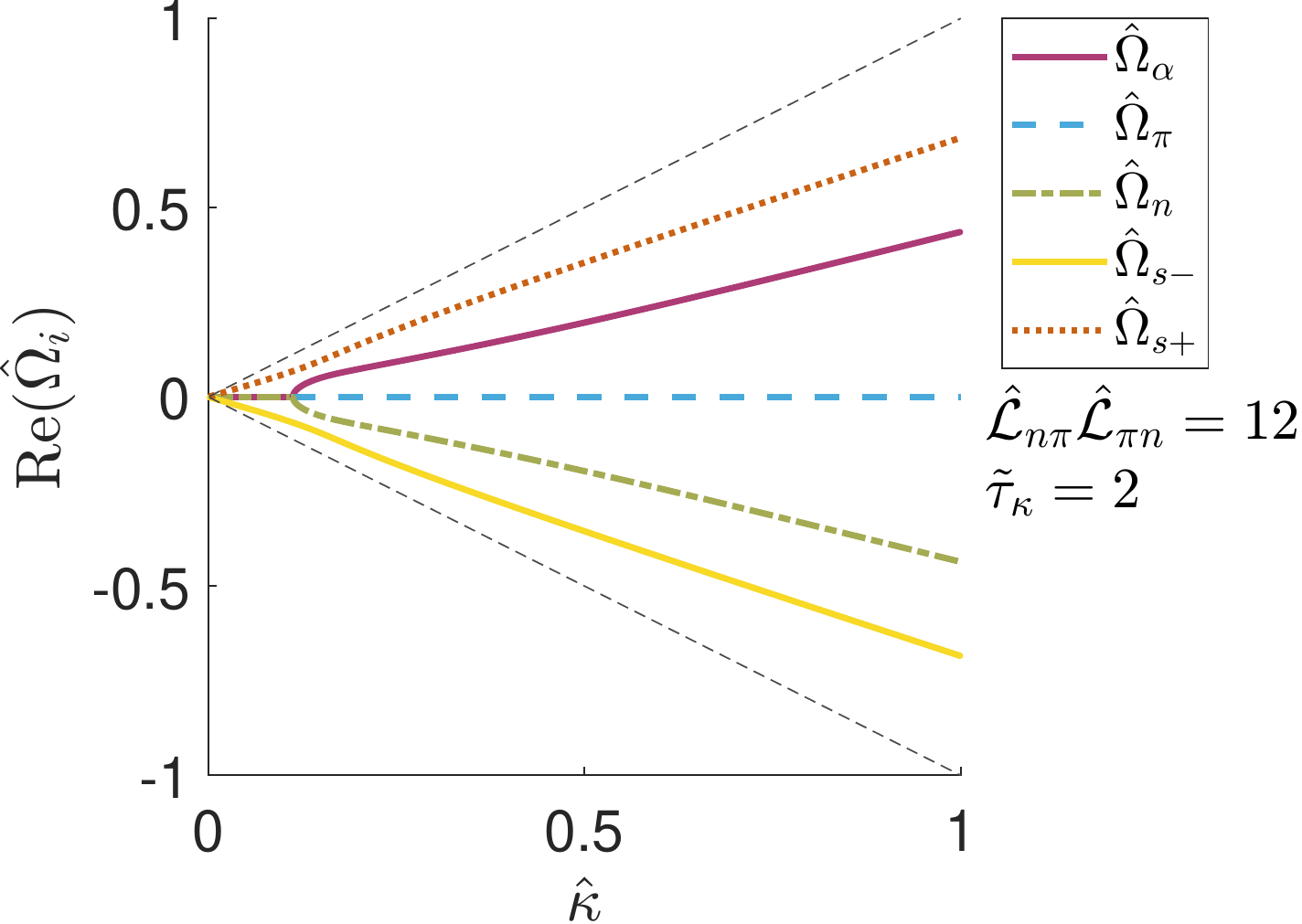}
	 \caption{The imaginary parts (upper panel) and real parts (lower panel) of the
	 five longitudinal modes, for $\hat{\tau}_\pi = 5$, $\hat{\tau}_n = 27/4 = 6.75$,  
	 $\tilde{\tau}_\kappa=2$, and 
	 $\hat{\mathcal{L}}_{n\pi} \hat{\mathcal{L}}_{\pi n} =12$ 
  (corresponding to point (e) in Fig.\ \ref{figure3}). Line modes are the same as in Fig.~\ref{figure1}.}
	 \label{figure8}
\end{figure}

\textit{Point (e)} is located at $(\tilde{\tau}_\kappa, \hat{\mathcal{L}}_{n\pi} \hat{\mathcal{L}}_{\pi n}) = (2,12)$ 
and the
corresponding dispersion relations are shown in Fig.\ \ref{figure8}. At this point, the shear mode is again identified with the stable, nonpropagating mode
$\hat{\Omega}_0 (\hat{\kappa})$, while all other modes are propagating. Only the two sound modes are unstable modes, while the charge-transport and charge diffusion modes are stable, since we are in a region in Fig.\ \ref{figure3} where condition (iii.1) is violated while (iii.2) is fulfilled.

\section{Conclusion and Outlook}
\label{sec:V}

In this work, we have studied the stability of relativistic second-order dissipative
fluid dynamics in the linear regime. We have extended the investigations of Ref.~\cite{Brito:2020nou} to the case of a nonzero
background charge. We have found that the transverse modes are not influenced by a nonzero
background charge at all. For the longitudinal modes of a system consisting of noninteracting, massless, 
classical particles, the only effect of a nonzero background charge 
is that the transport coefficient $\hat{\tau}_\kappa$ appearing in the dispersion relations
is replaced by $\tilde{\tau}_\kappa$, cf.\ Eq.\
(\ref{tildetaukappa}). Varying the background net-charge from 0 to $\pm \infty$, $\tilde{\tau}_\kappa$
changes from $\hat{\tau}_\kappa$ to $4 \hat{\tau}_\kappa$.

Further extending the analysis of Ref.~\cite{Brito:2020nou}, we have identified different regions 
in the $\tilde{\tau}_\kappa - \hat{\mathcal{L}}_{n\pi} \hat{\mathcal{L}}_{\pi n}$ plane.
For very large negative values of $\hat{\mathcal{L}}_{n\pi} \hat{\mathcal{L}}_{\pi n}$, the system
becomes acausal, which implies an instability in a moving background, in
agreement with the results of Ref.~\cite{Brito:2020nou}.
Increasing $\hat{\mathcal{L}}_{n\pi} \hat{\mathcal{L}}_{\pi n}$, we have found a stable and causal region,
which is further subdivided in three parts. In particular, we 
found a narrow band in the $\tilde{\tau}_\kappa - \hat{\mathcal{L}}_{n\pi} \hat{\mathcal{L}}_{\pi n}$ plane
where the dispersion relations exhibit a peculiar behavior. There is a range of $\hat{\kappa}$ values where
the imaginary parts of the shear and the charge-diffusion mode become degenerate, while their real
parts become nonzero. Further increasing $\hat{\kappa}$, the shear mode 
again becomes a nonpropagating non-hydrodynamic
mode, while the imaginary parts of the charge-transport and charge-diffusion modes become degenerate, indicating
that they develop into propagating modes.
We also investigated the region of positive values of $\hat{\mathcal{L}}_{n\pi} \hat{\mathcal{L}}_{\pi n}$
bounded from below by the conditions (\ref{cond_I}) and (\ref{cond_II}), 
where the system remains causal, but exhibits unstable modes. These conditions correspond to Eqs.\ (129) and (130) of 
Ref.~\cite{Brito:2020nou}, where they were not further investigated. This region can again be subdivided into five
parts characterized by a qualitatively different behavior of the various dispersion relations. We have studied
these analytically in the limit
of vanishing and large wave numbers and confirmed the results by an explicit numerical solution of Eq.\ (\ref{eq:60}).

This work can be extended along several lines. For instance, one could study a gas of massive particles and
also include the effect of the bulk pressure. In this case, however, one has to consider
the system (\ref{long_modes}) of longitudinal perturbations, which features an additional equation as 
compared to the simpler case (\ref{long_modes_ur}), where
the bulk pressure was neglected and the particles were assumed to be massless and classical.
A further line of extension is to include the effect of electromagnetic fields, e.g., based on the
equations of motion derived in Refs.~\cite{Denicol:2018rbw,Denicol:2019iyh}.
Preliminary work in this direction neglecting, however, the effects of charge transport and diffusion was already done
in Ref.\ \cite{Biswas:2020rps}.

\section*{Acknowledgments}
The authors acknowledge interesting discussions with V.\ Ambrus,
C.V.\ Brito, G.S.\ Denicol, S.\ Pu, and V.\ Roy.
This work is supported by the
Deutsche Forschungsgemeinschaft (DFG, German Research Foundation)
through the Collaborative Research Center TransRegio
CRC-TR 211 ``Strong- interaction matter under extreme conditions'' -- project number
315477589 -- TRR 211. The work of M.M.\ and D.H.R.\ is supported 
by the State of Hesse within the Research Cluster
ELEMENTS (Project ID 500/10.006).\\ ~~\\

\bibliography{Paper_net_charge_nonzero_301222}% Produces the bibliography via BibTeX.

\end{document}